\documentclass[twocolumn,superscriptaddress,floatfix,preprintnumbers, nofootinbib,hyperref]{revtex4-2} 
\pdfoutput=1
\usepackage[colorlinks=true,breaklinks=true]{hyperref}
\usepackage[normalem]{ulem}
\usepackage[utf8]{inputenc}
\hypersetup{allcolors=[rgb]{0.0 0.0 0.6},linkcolor=[rgb]{0.75 0.05 0.05}}
\usepackage{amsmath,amssymb}
\usepackage{epsfig}  
\usepackage{graphicx}   
\usepackage{slashed}       
\usepackage{tikz}
\usepackage{amsmath}
\usepackage{tikz-feynman}
\usepackage{url}
\usepackage{color}
\usepackage{multirow}
\usepackage{comment}
\usepackage{amssymb}
\usepackage{orcidlink}
\usepackage[version=3]{mhchem}
\DeclareUnicodeCharacter{2212}{-}

\DeclareMathOperator{\cm}{cm}
\DeclareMathOperator{\GeV}{GeV}

\DeclareMathOperator{\MeV}{MeV}

\DeclareMathOperator{\s}{s}

\DeclareMathOperator{\km}{km}

\DeclareMathOperator{\kpc}{kpc}
\DeclareMathOperator{\g}{g}

\DeclareMathOperator{\Mpc}{Mpc}

\newcommand{\gag}{\ensuremath{g_{a\gamma}}}

\allowdisplaybreaks

\begin{document}

\title{Supernova limits on `QCD axion-like particles'}

\author{Alessandro~Lella~\orcidlink{0000-0002-3266-3154}}
\email{alessandro.lella@ba.infn.it}
\affiliation{Dipartimento Interateneo di Fisica  ``Michelangelo Merlin,'' Via Amendola 173, 70126 Bari, Italy}
\affiliation{Istituto Nazionale di Fisica Nucleare - Sezione di Bari, Via Orabona 4, 70126 Bari, Italy}%

\author{Eike Ravensburg \orcidlink{0000-0001-5827-9479}}
\email{ravensburg@cp3.sdu.dk}
\affiliation{The Oskar Klein Centre, Department of Physics, Stockholm University, Stockholm 106 91, Sweden}
\affiliation{CP3-Origins, University of Southern Denmark,
Campusvej 55, 5230 Odense M, Denmark}

\author{Pierluca~Carenza~\orcidlink{0000-0002-8410-0345}}\email{pierluca.carenza@fysik.su.se}
\affiliation{The Oskar Klein Centre, Department of Physics, Stockholm University, Stockholm 106 91, Sweden}

\author{M.C.~David~Marsh~\orcidlink{0000-0001-7271-4115}}\email{david.marsh@fysik.su.se}
\affiliation{The Oskar Klein Centre, Department of Physics, Stockholm University, Stockholm 106 91, Sweden}

\date{\today}
\smallskip

\begin{abstract}
    In this paper, we explore the phenomenology of massive Axion-Like Particles (ALPs) coupled to quarks and gluons, dubbed `QCD ALPs', with an emphasis on the associated low-energy observables. ALPs coupled to gluons and quarks not only induce nuclear interactions at scales below the QCD-scale, relevant for ALP production in supernovae (SNe), but naturally also couple to photons similarly to the QCD-axion. We discuss the link between the high-energy formulation of ALP theories and their effective couplings with nucleons and photons. The induced photon coupling allows ALPs with masses $m_a\gtrsim1\,\MeV$ to efficiently decay into photons, and astrophysical observables severely constrain the ALP parameter space. We show that a combination of arguments related to SN events rule out ALP-nucleon couplings down to $g_{aN}\gtrsim 10^{-11}- 10^{-10}$ for $m_a\gtrsim1 \MeV$ -- a region of the parameter space that was hitherto unconstrained.
\end{abstract}
\maketitle

\section{Introduction}

Several extensions of the Standard Model (SM) predict the existence of pseudo-scalar particles still unobserved (see e.g.~\cite{Jaeckel:2010ni,DiLuzio:2020wdo} for recent reviews). The most relevant example is the axion~\cite{Weinberg:1977ma,Wilczek:1977pj}, which is introduced to solve the strong-CP problem of Quantum Chromodynamics (QCD) via the Peccei-Quinn (PQ) mechanism~\cite{Peccei:1977hh,Peccei:1977ur}. Nevertheless, several pseudo-scalar fields similar to axions naturally emerge as pseudo-Nambu-Goldstone bosons of global symmetries that are broken at some high-energy scale $f_a$, without solving in general the strong-CP problem. To distinguish them from the QCD axion, they are dubbed Axion-Like Particles (ALPs). ALP properties reflect the high-energy theory from which they originated. For instance, ALPs with a mass spectrum stretching over a large range of scales are predicted by string theory and feature a very interesting phenomenology~\cite{Svrcek:2006yi,Cicoli:2012sz,Halverson:2019kna}.  It has long been realized that axions and ALPs can comprise the dark matter~\cite{Preskill:1982cy,Abbott:1982af,Dine:1982ah}, and the past few years have witnessed a flurry of activity in this field, including new experimental proposals that explore deep into the fundamental parameter space~\cite{Lawson:2019brd,BREAD:2021tpx,DMRadio:2022pkf}.

At low energies, the ALP phenomenology is determined by the effective couplings to photons and matter fields
\begin{equation}
\label{eq:interactions}
\mathcal{L} \supset \frac{1}{4}g_{a\gamma} a\, F_{\mu\nu} \tilde F^{\mu\nu} +\sum_{N} g_{aN}  \frac{\partial_\mu a}{2m_{N}} \bar{N} \gamma^\mu \gamma_5 N +\frac{m_{a}^{2}}{2}a^{2} \, ,
\end{equation}
where $g_{a\gamma}$ is the ALP-photon coupling, $F_{\mu\nu}$ is the electromagnetic field strength tensor, $\tilde F^{\mu\nu}$ is its dual, $N = p,n$ represents nucleons with masses $m_{N}$, $g_{aN}$ are the ALP-nucleon couplings and $m_{a}$ is the ALP mass. Contrary to QCD-axion models, in this case $m_a$ is a free parameter not related to ALP couplings.

Many of the theoretical and experimental efforts are focused on the ALP-photon coupling $\gag$ in the first term of Eq.~\eqref{eq:interactions}. In presence of an external magnetic field, the ALP-photon interaction leads to the phenomenon of ALP-photon conversion~\cite{Raffelt:1987im}. This effect is widely used by several ongoing and upcoming ALP experiments~\cite{Irastorza:2018dyq,DiLuzio:2020wdo,Sikivie:2020zpn} to probe the existence of axions and ALPs. 
The same coupling would also allow for the ALP production in stellar plasmas via the Primakoff process, i.e.~ALP-photon conversion in the electromagnetic field generated by the plasma~\cite{Raffelt:1985nk}. Thus, observations of the Sun, globular clusters and supernovae (SNe) induce severe constraints on the ALP-photon coupling~\cite{Caputo:2024oqc,DiLuzio:2021ysg}. Moreover, the ALP-photon coupling leads to important signatures in astrophysical photon spectra, from X-ray to PeV energies~\cite{Wouters:2013hua, Berg:2016ese,Reynolds:2019uqt,Reynes:2021bpe,Fermi-LAT:2016nkz,Davies:2022wvj,Li:2021gxs,Meyer:2016wrm,Meyer:2020vzy,Jacobsen:2022swa,MAGIC:2024arq,Mastrototaro:2022kpt,Calore:2023srn}.\\
The ALP-nucleon couplings, $g_{aN}$, in Eq.~\eqref{eq:interactions} also lead to ALP production in different stellar systems, e.g.~via $NN$-bremsstrahlung  and pion conversion in SNe~\cite{Burrows:1988ah,Burrows:1990pk,Carenza:2019pxu,Carenza:2021ebx,Fischer:2021jfm,Lella:2022uwi,Lella:2023bfb,Carenza:2023lci}. In addition, experimental techniques sensitive to nuclear couplings have been recently proposed~\cite{Arvanitaki:2014dfa,Crescini:2017uxs,JacksonKimball:2017elr}. However, as we argue in this paper, the ALP's low-energy couplings to both nucleons and photons originate from a common high-energy theory and are not completely independent. In particular, an ALP coupled to nucleons naturally also couples to photons, with strong implications for the resulting phenomenology. The low-energy effective Lagrangian in Eq.~\eqref{eq:interactions} provides an effective description of the ALP interactions with the Standard Model at energies below the QCD confinement scale. At higher energies, such interactions are expected to be generated by couplings to the fundamental degrees of freedom, i.e.~the quarks and gauge fields \cite{Georgi:1986kr,Georgi:1986df,Peccei:1988ci,GrillidiCortona:2015jxo,Choi:2021ign}. In this paper, we consider \emph{`QCD ALPs'} with interactions to quarks and gluons given by
\begin{equation}
\begin{aligned}
\mathcal{L}_{\rm aQCD} &= c_{g} \frac{g_s^{2}}{32\pi^{2}} \frac{a}{f_a} 
G^a_{\mu\nu} \tilde G^{a\mu\nu} +\sum_{q} c_{q} \frac{\partial_\mu a}{2 f_a} \bar q \gamma^\mu \gamma_5 q\\
&\quad + \frac{(m_{a,0})^{2}}{2}a^{2}\,,
\end{aligned}
\label{eq:LaaboveQCD}
\end{equation}
where $g_{s}$ is the coupling constant of QCD, $f_a$ is the axion decay constant,  $G^a_{\mu\nu}$ is the gluon field strength tensor, $\tilde G^a_{\mu\nu}=\frac{1}{2}\,\epsilon_{\mu\nu\rho\sigma}G^{a\rho\sigma}$ its dual, $c_{g}$ and $c_{q}$ are model-dependent constants, and $q=u,d,s,c,t,b$ runs over the quark species. The QCD ALP is distinguished from the QCD axion by also coupling to a hidden, i.e.~non-Standard Model, gauge sector that confines at a high scale and is assumed to generate a mass, $m_{a,0}$, to the ALP. We will be interested in the case where the ALP is stabilized by $m_{a,0}$ at a sufficiently high scale to decouple from the dynamics resolving the strong CP-problem, cf.~Appendix \ref{sec:axionAndALP} for more details. Here, we focus on the phenomenology resulting from the coupling of the massive ALP to quark and gluons in Eq.~\eqref{eq:LaaboveQCD}, resulting in an ``irreducible'' coupling of the ALP to photons in addition to ALP-nucleon couplings. We will focus on the parameter space in which ALPs are significantly produced during a SN explosion by means of nuclear processes (cf.~Ref.~\cite{Lella:2023bfb} for recent developments). We point out that, due to the induced photon coupling, the massive ALPs produced in SNe can rapidly decay into photon pairs giving rise to directly or indirectly observable signatures that can be compared to data, such as observations of gamma-rays near Earth or an alteration to the typical explosion energies of core-collapse SNe.
Specifically, we focus on the range of ALP masses $ 1 \MeV \lesssim m_a \lesssim 700\MeV $ where ALP radiative decays are efficient, leading to the most stringent limits.
The paper is organized as follows. In Sec.~\ref{sec:ALPinteractions} we present QCD ALP interactions below the QCD confinement scale, illustrating why both couplings to nucleons and photons emerge naturally in our framework. In Sec.~\ref{sec:ALPsfromSNe} we describe the main ALP production channels in a SN core. In Sec.~\ref{sec:ALPdecays} we discuss the main decay channels for ALPs with masses $m_a\gtrsim10\,\MeV$ and the related phenomenology. In particular in Sec.~\ref{sec:EnergyDeposition} we study the scenario in which they can deposit energy in the SN envelope, in Sec.~\ref{sec:decayBound} the possibility they have induced a gamma-ray burst in coincidence to SN 1987A and in Sec.~\ref{sec:DSNALPB} the eventuality they have given rise to a diffuse background from all past SNe.
In Sec.~\ref{sec:results} we analyze our results and discuss the bounds introduced for the benchmark cases considered. Finally, in Sec.~\ref{sec:conclusions} we conclude.

\section{ALP interactions with nucleons and photons}
\label{sec:ALPinteractions}

The interaction Lagrangian $\mathcal{L}_{\rm aQCD}$, described in Eq.~\eqref{eq:LaaboveQCD}, induces ALP couplings with baryons and mesons at energies below the QCD confinement scale~\cite{Georgi:1986kr,Georgi:1986df,Peccei:1988ci,GrillidiCortona:2015jxo,Choi:2021ign}. In analogy to the QCD axion case, the nuclear interaction Lagrangian can be derived in the context of Heavy Baryon Chiral Perturbation Theory (HBChPT)~\cite{Ho:2022oaw}, valid for non-relativistic baryons. Starting from the couplings in Eq.~\eqref{eq:LaaboveQCD} defined at energy scales $\mu \gtrsim 1\,\GeV$, the relevant low-energy ALP interactions with nucleons, pions and baryonic resonances read~\cite{Ho:2022oaw,Choi:2021ign,GrillidiCortona:2015jxo}:
\begin{equation}
    \begin{split}
        \mathcal{L}_{\rm{nuc}}&=\frac{\partial^\mu a}{2f_a}\Bigg[C_{p}\Bar{p}\gamma^\mu\gamma_5p+C_{n}\Bar{n}\gamma^\mu\gamma_5n\\
        &+\frac{C_{a\pi N}}{f_\pi}(i\pi^+\Bar{p}\gamma^\mu n-i\pi^-\Bar{n}\gamma^\mu p)  \\
        &+C_{aN\Delta}\left(\Bar{p}\,\Delta^+_\mu+\overline{\Delta^+_\mu}\,p+\Bar{n}\,\Delta^0_\mu+\overline{\Delta^0_\mu}\,n\right)\Bigg],
    \end{split}
\label{eq:NuclearInteractions}
\end{equation}
where $f_{\pi}=92.4~\MeV$ is the pion decay constant, $C_p$ and $C_n$ are ALP couplings to protons and neutrons, respectively. 
They also determine the couplings to pions and the $\Delta$-resonances $C_{a\pi N}$ and $C_{aN\Delta}$ 
as~\cite{Ho:2022oaw} 
\begin{equation}
    C_{a\pi N}=\frac{(C_{p}-C_{n})}{\sqrt{2}\,g_{A}}\,,\,\,\,\,\,C_{aN\Delta}=-\frac{\sqrt{3}}{2}(C_{p}-C_{n})\,,
\end{equation}
where $g_{A}=1.28$~\cite{ParticleDataGroup:2022pth} is the axial coupling.
Furthermore, the low-energy parameters $C_p$ and $C_n$ are related to $c_g$ and $c_q$, as they appear in Eq.~\eqref{eq:LaaboveQCD}, by the following relations~\cite{GrillidiCortona:2015jxo}:
\begin{equation}
    \begin{split}
    C_p(c_g,c_u,c_d)=&-0.47\,c_g+0.88c_{u}-0.39c_{d}-0.038c_{s}\\
    &-0.012c_{c}-0.009c_{b}-0.0035c_{t}\,,\\
    C_n(c_g,c_u,c_d)=&-0.02\,c_g+0.88c_{d}-0.39c_{u}-0.038c_{s}\\
    &-0.012c_{c}-0.009c_{b}-0.0035c_{t}\,.
    \end{split}
    \label{eq:coupl}
\end{equation}
In particular, these expressions were computed in Ref.~\cite{GrillidiCortona:2015jxo} by assuming $f_a = 10^{12} \, \GeV$. However, the variation in the coefficients due to changes in the matching scale is negligible compared to the theoretical uncertainties from lattice simulations~\cite{GrillidiCortona:2015jxo}. Neither has any significant impact on the phenomenology or the uncertainty on the bound.
Eq.~\eqref{eq:coupl} suggests that the contribution from $s$ and heavier quarks is suppressed at least by 1 order of magnitude with respect to the lighter quarks $u,d$. Therefore, we neglect the contribution from heavier quarks and will only consider ALP couplings to gluons and 
to light quarks ($u,d$). 

The Lagrangian of Eq.~\eqref{eq:LaaboveQCD} does not include a tree-level ALP-photon coupling, but such a coupling is generated in the low-energy theory through fermion loops and axion-pion mixing, as described in Ref.~\cite{Bauer:2017ris}. Thus, Eq.~\eqref{eq:NuclearInteractions}
should be complemented by the effective Lagrangian term 
\begin{equation}
    \mathcal{L}_{a\gamma\gamma}=-\frac{1}{4}\,g_{a\gamma}\, a F_{\mu\nu}\Tilde{F}^{\mu\nu}\,,
\end{equation}
where $g_{a\gamma}=\alpha_{\rm em} C_\gamma/(2\pi f_a)$, with the fine-structure constant $\alpha_{\rm em}$, and~\cite{Bauer:2017ris,Bauer:2021mvw}
\begin{equation}
    \begin{split}
        C_{\gamma}(c_g,c_u,c_d)&=-1.92\,c_g
        \\
        &\quad\quad-\frac{m_a^2}{m_\pi^2-m_a^2}\Bigg[c_g\frac{m_d-m_u}{m_d+m_u}+(c_u-c_d)\Bigg]\,,
        \label{eq:C_gamma}
    \end{split}
\end{equation}
which holds for $m_a\lesssim 1\,\GeV$ and away from the strong mixing regime $|m_\pi^2 - m_{a,0}^2| \gg m_\pi^2 f_\pi/f_a$ \cite{Bauer:2020jbp}. In this expression, $m_u$ and $m_d$ are the masses of the light quarks whose ratio is measured from lattice estimates $m_u/m_d=0.48$~\cite{deDivitiis:2013xla,MILC:2015ypt,Horsley:2015eaa,GrillidiCortona:2015jxo}. We highlight that also in this case the contribution from heavier quarks can be safely neglected, as discussed in Refs.~\cite{Bauer:2021mvw,Bauer:2017ris}. In particular, the $s$ quark contribution is suppressed by factors $\mathcal{O}(m_{u,d}/m_s)$, while additional terms from $q=c,b,t$ quarks are suppressed by terms $\mathcal{O}(m_a^2/m_q^2)$.

Finally, the ALP-gluon coupling introduces loop corrections also to the ALP mass $m_a$~\cite{Bauer:2017ris}. Considering only Eq.~\eqref{eq:LaaboveQCD}, and no QCD axion state, this leads to the corrected ALP mass
\begin{equation}
    m_{a}^{2}=c_{g}^{2}m_{\rm QCD}^{2} + \left[1 + \mathcal{O}\left(\frac{f_\pi^2}{f_a^2}\right)\right] m_{a,0}^{2}\,,
    \label{eq:ALPmass}
\end{equation}
where $m_{\rm QCD}=5.70\,{\rm \mu eV}\left(10^{12}\GeV/f_{a}\right)$ is the mass term induced by QCD effects~\cite{GrillidiCortona:2015jxo,DiLuzio:2020wdo}. Clearly, including the QCD axion leads to mixing, as we discuss in more detail in Appendix \ref{sec:axionAndALP} (cf.~\cite{Gavela:2023tzu}). For $m_{a,0}\gg m_{\rm QCD}$, the QCD ALP is the mass eigenstate with $m \simeq m_{a,0}$.

\subsection*{The induced photon coupling}
In this subsection, we describe how we will study and parametrize the ALPs coupled to nucleons and photons introduced above. Both, $\mathcal{L}_{\rm aQCD}$ and the resulting low-energy EFT described by $\mathcal{L}_{\rm nuc} + \mathcal{L}_{a\gamma\gamma}$ have 4 free parameters (ignoring the heavy quarks as justified above). In phenomenological studies, the ALP-nucleon or ALP-photon couplings, as well as the ALP mass, are usually employed as parameters for the theory. In this spirit, we can recast
Eq.~\eqref{eq:coupl}
to express the quark couplings in terms of the proton and neutron couplings:
\begin{equation}
    \begin{split}
    c_{u}&=0.68 c_{g}+0.63C_{n}+1.41C_{p}\,,\\
    c_{d}&=0.32 c_{g}+1.41 C_{n}+0.63 C_{p}\,.
    \label{eq:couplInverse}
    \end{split}
\end{equation}
Since there are three fundamental couplings and two nucleon couplings, the gluon coupling has to be left as a free parameter here; 
in general, an ALP coupled to nucleons is coupled to both gluons and quarks.

Since in Eq.~\eqref{eq:LaaboveQCD} the scaling of all $c_g$ and $c_q$ by the same factor is equivalent to a rescaling of $f_a$, we we can restrict to two cases:
\begin{itemize}
    \item \textit{$c_{g}=0$}. ALPs are decoupled from the gluon field and they only interact with light quarks. In this scenario, the induced coupling to photons reads
    \begin{equation}
        C_{\gamma}\simeq-0.79\frac{m_{a}^{2}}{m_{\pi}^{2}-m_{a}^{2}}\left(C_{p}-C_{n}\right)\,,
        \label{eq:Cgammacg0}
    \end{equation}
    which is strongly mass-dependent and sizable for ALP masses $m_a>\mathcal{O}(10)\MeV$.
    The coupling is suppressed for $C_{p}=C_{n}$, which (with $c_g=0$) corresponds to $c_u=c_d$. Such a cancellation may occur in DFSZ-like ALP models if the vacuum expectation values of the two additional Higgs fields are equal, but this cancellation is not generic. In the absence of tuning, we expect that $|C_{\gamma}|\gtrsim \mathcal{O}(10^{-1}) \left(\frac{m_{a}}{m_{\pi}}\right)^{2}$.
    \item \textit{$c_{g}=1$}. In this case, ALPs are coupled to both gluons and light quarks and the induced ALP-photon coupling can be written as
    \begin{equation}
        \begin{split}
        C_{\gamma}\simeq-1.92-\frac{m_{a}^{2}}{m_{\pi}^{2}-m_{a}^{2}}\Big[0.71+0.79\left(C_{p}-C_{n}\right)\Big]\,,
        \end{split}
    \label{eq:Cgammacg1}
    \end{equation}
 where the first term results from the `irreducible' infrared mixing with pions. Note that $C_\gamma$ is sizable independently of the ALP mass, $|C_{\gamma}|\sim\mathcal{O}(1)$.
\end{itemize}
We emphasize that the ALP-photon coupling emerges naturally in theories of ALPs that couple to quarks and gluons, and can only be avoided through tuning of the microscopic parameters.
This has important phenomenological implications for ALPs coupled to nucleons. In particular, phenomena related to the ALP emission from a nuclear medium should not be limited to only account for nuclear processes, but also the production and decay channels due to the induced ALP-photon coupling.

Starting from the expressions in Eq.~\eqref{eq:Cgammacg0} and Eq.~\eqref{eq:Cgammacg1}, it is possible to express the dimensionful ALP-photon coupling $g_{a\gamma}$ as a function of the ALP-nucleon coupling $g_{aN}=C_N\,m_N/f_a$. For the sake of simplicity, in the following, we will assume $C_n\simeq0$. This choice is useful in order to reduce the number of free parameters considered in the analysis. Under this assumption, the axion-photon coupling reads
\begin{equation}
\begin{split}
        c_{g}=0\,, \,\,\,\,\,\,\,\,\,   g_{a\gamma}&\simeq -9.7\times10^{-4}\,\frac{m_{a}^{2}}{m_{\pi}^{2}-m_{a}^{2}}\,g_{ap}\,\GeV^{-1}\,,\\
        c_{g}=1\,, \,\,\,\,\,\,\,\,\,   g_{a\gamma}&\simeq 
        -9.5\times10^{-4}\,
        g_{ap}\,\GeV^{-1} \\
        &\times \Bigg[\frac{1.53}{c_d-0.33} + \frac{c_d + 0.24}{c_d-0.33}\frac{m_{a}^{2}}{m_{\pi}^{2}-m_{a}^{2}}\Bigg]
    \label{eq:photon_coupling2}
    \end{split}
\end{equation}
where we used the condition $C_{n}\simeq0$ in order to fix $c_{u}$, but $c_d$ is still a free parameter of the assumed ALP model. For the discussion of our results, we will assume as benchmark values $c_d=0$ and $c_d=1$. Note that $c_d$ is not an independent parameter in the $c_g = 0$ case, where $C_n=0$ and $g_{ap}$ fully determine the two fundamental couplings to the light quarks. For $c_g = 1$ on the other hand, $c_d$ determines the relative strengths of the ALP couplings to gluons and protons. 

Moreover, the induced ALP-photon coupling shows a pole at $m_a = m_\pi$. As remarked below Eq.~\eqref{eq:C_gamma}, our results do not hold very close to the pole, i.e.~for ALP masses near the pion mass $m_\pi\simeq135\,\MeV$. In the range of couplings considered in this work, $g_{aN}\lesssim10^{-8}$, we have $f_a \gtrsim 10^{-8}\,\GeV^{-1}$ and Eqs.~\eqref{eq:photon_coupling2} hold for $|m_a - m_\pi|/m_\pi \gtrsim 10^{-9}$~\cite{DiLuzio:2024jip}. Thus, this criterion does not meaningfully limit the phenomenologically interesting parameter range.

\begin{figure}
\centering
    \includegraphics[width=1\columnwidth]{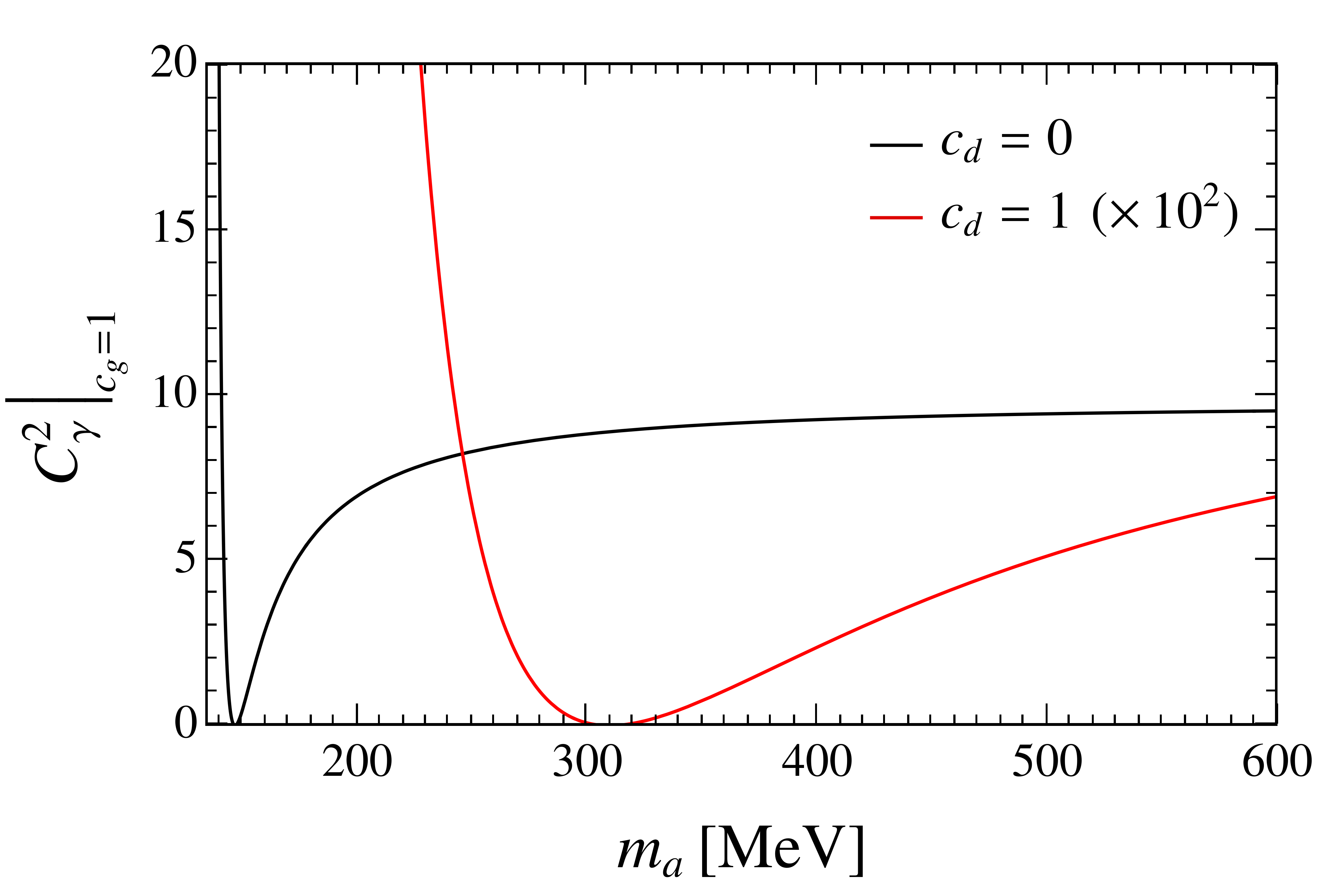}
    \caption{Behavior of $C_\gamma^2$ defined as in Eq.~\eqref{eq:C_gamma} in the $c_g=1$ scenario for ALP masses $m_a>m_\pi$. The black and red lines refer to the $c_d=0$ and $c_d=1$ cases, respectively. To better appreciate the presence of the minimum, the $c_d=1$ lineshape has been rescaled by a factor $100$.}

    \label{fig:Cgamma}
\end{figure}

Furthermore, in the $c_g=1$ scenario, the induced ALP-photon coupling is suppressed for some special values of the ALP mass in both benchmark cases considered. In the mass range $m_a\gtrsim m_\pi$, this eventuality occurs when the second term in the parentheses of Eq.~\eqref{eq:photon_coupling2} cancels the first one. This is equivalent to having the reciprocal cancellation of the two terms in Eq.~\eqref{eq:C_gamma}. In Fig.~\ref{fig:Cgamma}, we show that $C_\gamma$ vanishes at $m_a \simeq 147\,\MeV$ and $m_a \simeq 310\,\MeV$ for the $c_d=0$ and $c_d=1$ cases, respectively.
Important for this work, ALPs with those masses cannot decay into photon pairs, and hence, the observable signatures and the resulting constraints which are driven by these decays have a strong mass dependence near these parameter values.

\section{ALP production from supernovae}
\label{sec:ALPsfromSNe}

Core-collapse SNe are unique astrophysical laboratories to search for ALPs coupled to nuclear matter. Because of the extreme conditions of temperature and density expected in the inner regions of the proto-neutron star (PNS), ALPs could be copiously produced by means of nuclear processes. In recent years, the characterization of the ALP emission from the hot and dense nuclear medium of an exploding SN core has been revealed to be more complex than was originally thought in some pioneering works at the end of the 80's~\cite{Carena:1988kr,Brinkmann:1988vi,Raffelt:1993ix,1996slfp.book.....R,Turner:1991ax,Keil:1996ju}. \newline
The first channel for axion production in the hot and dense SN nuclear medium is nucleon-nucleon ($NN$) bremsstrahlung $N+N\rightarrow N+N+a$, which was believed to be the dominant process for many years. The state-of-the-art calculation for the emission rate associated to this process is illustrated in Ref.~\cite{Carenza:2019pxu} and accounts for corrections beyond the usual One-Pion-Exchange (OPE) approximation of the nuclear interaction potential~\cite{Ericson:1988wr}, effective nucleon masses in the SN core and multiple scattering effects~\cite{Raffelt:1991pw,Janka:1995ir}. However, starting from the seminal idea of Ref.~\cite{Carenza:2020cis}, it has been realized that if the density of negatively charged pions in the core is large enough~\cite{Fore:2019wib}, the contribution due to pionic Compton-like processes $\pi+N \rightarrow a+N$ could significantly enhance the ALP production and even be dominant with respect to $NN$ bremsstrahlung. Therefore, the contributions coming from both processes have to be taken into account. Furthermore, for ALP-nucleon couplings $g_{aN}\lesssim10^{-8}$, SN ALPs are in the \textit{free-streaming} regime, in which reabsorption effects due to inverse nuclear processes can be neglected~\cite{Lella:2023bfb}. The complete expressions for ALP emission spectra, including finite mass effects for $m_a\gtrsim10\,\MeV$ are provided in Refs.~\cite{Lella:2022uwi,Carenza:2023lci}. Remarkably, if pions are present in the SN core, the emission spectrum of SN ALPs coupled to nucleons shows a peculiar bimodal shape due to the different energy ranges in which the two different production mechanisms are efficient~\cite{Lella:2023bfb}.  In particular, bremsstrahlung and pion conversion spectra peak at energies $E_a\sim50\MeV$ and $E_a\sim200 \MeV$, respectively.

Naively, in the non-degenerate regime for nucleons and pions, the ALP emissivities , i.e. the energy released in ALPs per unit volume and unit time, can be simply estimated as~\cite{Caputo:2024oqc}
\begin{equation}
    \begin{split}
        Q_a^{NN}&=g_{ap}^2\,\rho\,\frac{T^4}{4\,\pi^2\,m_N^2}\,F\,,\\
        Q_a^{\pi N}&=\frac{15}{\pi^3}\frac{g_{ap}^2}{m_N^2}\left(\frac{g_A}{f_\pi}\right)^2\,n_p\,z_\pi T^6\,,
    \end{split}
\label{Eq:rates}
\end{equation}
for $NN$-bremsstrahlung and pion conversion, respectively. In particular, in these expressions we have introduced the local SN temperature $T$ and density $\rho$, the nucleon mass $m_N$, the proton number density $n_p$, the pion fugacity $z_\pi$, the pion decay constant $f_\pi=92.4\,\MeV$ and the axial coupling $g_A=1.27$~\cite{ParticleDataGroup:2022pth}, while $F$ is an $\mathcal{O}(1)$ numerical factor.

In our analysis we will employ as reference SN model the 1D spherical symmetric {\tt GARCHING} group's SN model SFHo-s18.8 provided in~\cite{SNarchive}, starting with a stellar progenitor with mass $18.8~M_\odot$~\cite{Sukhbold:2017cnt} and based on the neutrino-hydrodynamics code {\tt PROMETHEUS-VERTEX}~\cite{Rampp:2002bq}. Our benchmark model, already used in previous analyses (see, e.g., Refs.~\cite{2020PhRvL.125e1104B,Caputo:2021rux,Caputo:2022mah,Fiorillo:2023frv}), leads to a neutron star with baryonic mass $1.351~M_\odot$ and gravitational mass $1.241~M_\odot$. We highlight that the presence of pions inside the SN core is still under debate (see~\cite{Fore:2023gwv} for a recent analysis). Therefore, we estimated the pion chemical potential and a pion abundance on top of the {\tt GARCHING} group's SN simulation, by employing the procedure in Ref.~\cite{Fischer:2021jfm}, including the pion-nucleon interaction as described in Ref.~\cite{Fore:2019wib}. 

As suggested by the strong dependence on the SN temperature shown by the ALP emivissivities in Eq.~\eqref{Eq:rates}, the ALP production in the PNS nuclear medium is very sensitive to SN conditions. However, we point out that the benchmark SN model employed in this work is characterized by SN peak temperatures $T\simeq35\,\MeV$ and peak densities $\rho\simeq3\times10^{14}\g \cm^{-3}$ and it is among the coldest model available in the {\tt GARCHING} group archive. In particular, the employed SFHo-s18.8 SN profile coincide to the ``cold'' model of Ref.~\cite{Caputo:2021rux}, where the authors argued that this SN profile typically leads to lower ALP emission rates compared to other models. Therefore, limits derived by employing this model have to be considered conservative.\\
As discussed in Ref.~\cite{Lella:2023bfb}, a possible source of uncertainty in the ALP free-streaming regime is related to the presence of a relatively high fraction of pions in the SN core $Y_\pi\sim\mathcal{O}(10^{-2})$, which is still under debate. If the pion abundance is suppressed, the pion conversion contribution is reduced and SN bounds could result less stringent. However, following the discussion of Ref.~\cite{Lella:2023bfb}, in the low mass limit we expect constraints to be relaxed by no more than a factor $\sim2$, which is smaller than the typical uncertainty due to, e.g., possibly higher temperatures or densities. Moreover, we expect that only lower ALP masses can be probed and constraints could vanish for ALP masses $m_a\gtrsim200\,\MeV$, where ALP production via bremsstrahlung becomes inefficient.

As discussed before, together with nuclear processes, also the ALP-photon interaction contributes to the number of ALPs produced in the SN via the 
Primakoff process and photon coalescence~\cite{Lucente:2020whw}. However, a simple estimate makes it clear that these production channels are strongly suppressed with respect to bremsstrahlung and pionic processes in the ALP free-streaming regime, in which we will develop our analysis. Indeed, even for $g_{ap}\sim10^{-8}$, the induced ALP photon coupling is $g_{a\gamma}\sim\mathcal{O}(10^{-11})\GeV^{-1}$, which is well below the SN cooling bound placed in Ref.~\cite{Lucente:2020whw} $g_{a\gamma}<6\times10^{-9}\,\GeV^{-1}$, obtained by considering the Primakoff process and photon coalescence for a tree-level ALP-photon coupling. Thus, in this coupling range, the ALP luminosity via ALP-photon interactions is much smaller than the neutrino luminosity $L_\nu$. On the other hand, at $g_{ap}\sim10^{-8}$ production via nuclear processes leads to an axion luminosity $L_a\sim100\,L_\nu$ (see Fig.~1 of Ref.~\cite{Lella:2023bfb}). Therefore, axion production by the ALP-photon coupling is many orders of magnitude smaller than nuclear production and can be safely neglected in our study. This also means that the ALP-photon coupling has no impact on the cooling argument, and the constraint derived in Ref.~\cite{Lella:2023bfb} is not modified.

\section{ALP decays}
\label{sec:ALPdecays}

Even though the ALP-photon coupling is inefficient as ALP production channel, it is responsible for ALP decays into photon pairs, which have very important phenomenological implications. The decay rate in the rest frame of an ALP with mass $m_a$ is given by~\cite{Raffelt:2006rj}
\begin{equation} \Gamma_{a\gamma\gamma}=g_{a\gamma}^2\frac{m_a^3}{64\pi}\,,
\label{eq:decayrate}
\end{equation}
corresponding to a decay length in a frame in which the ALP has energy $\omega_a$~\cite{Calore:2021klc} 
\begin{equation}
\begin{split}
\label{eq:lambdaGamma}
        \lambda_\gamma=\frac{\gamma_a \beta_a}{\Gamma_{a\gamma\gamma}} &\simeq 0.13\kpc\left(\frac{p_{a}}{m_a}\right)\left(\frac{m_{a}}{10\MeV}\right)^{-3}\\
        &\quad \times\left(\frac{g_{a\gamma}}{10^{-13}\GeV^{-1}}\right)^{-2}\,,
\end{split}
\end{equation}
where $\gamma_a$ is the Lorentz factor, $\beta_a=\sqrt{1-(m_a/\omega_a)^2}$ the relativistic velocity and $p_a$ is the ALP momentum.

We note that the ALP-photon coupling can also induce an  ALP-electron interaction at the one-loop level~\cite{Bauer:2017ris}. In this case, the effective constant encoding the electron coupling is given by
\begin{equation}
    c_{e}=-\frac{3\alpha_{\rm em}^{2}}{8\pi^{2}}C_{\gamma}K\,,
\end{equation}
in which $K=\log\left({f_a^2}/{m_e^2}\right)+\delta_1+g(m_a)$ and the renormalization scheme dependent constant $\delta_1$, as well as the function $g(m_a)$, are provided in Ref.~\cite{Bauer:2017ris}. In particular, $K$ is an $\mathcal{O}(10)$ factor at scales $f_a\sim10^9\,\GeV$ and masses $m_a\gtrsim10\,\MeV$.
Furthermore, the induced ALP-electron coupling $g_{ae}=m_e c_e/f_a$ reads 
\begin{equation}
    g_{ae}=\frac{3\alpha_{\rm em}\,m_{e}}{4\pi}K\,g_{a\gamma}\sim 8.7\times10^{-7}K\,\left(\frac{g_{a\gamma}}{\GeV^{-1}}\right)\,.
\end{equation}
Introducing the ALP decay length for electron-positron decays~\cite{Jaeckel:2017tud,Altmann:1995bw}
\begin{equation}
    \begin{split}
        \lambda_{e}=&\frac{8\pi}{g_{ae}^2m_a}\frac{\omega_a}{m_a}
    \sqrt{\frac{1-m_a^2/\omega_a^2}{1-4m_\ell^2/m_a^2}} = \\ 
    =&\,0.016\,{\rm kpc}\left(\frac{p_{a}}{m_a}\right)\,\left( \frac{m_a}{10\,{\rm MeV}} \right)^{-1}\,\sqrt{1-\frac{4m_e^2}{m_a^2}}\\
    &\times\left( \frac{g_{ae}}{10^{-15}} \right)^{-2}\,,
    \end{split}
\end{equation}
we can estimate the branching ratio for decays into electron-positron pairs as 
\begin{equation}
    \begin{split}
        &\mathrm{BR}(a\rightarrow e^+e^-)=\frac{\lambda_e^{-1}}{\lambda_e^{-1}+\lambda_\gamma^{-1}}\simeq\frac{\lambda_e^{-1}}{\lambda_\gamma^{-1}}\\
        &\simeq6.1\times10^{-6}\,\left(\frac{m_a}{10\,\MeV}\right)^{-2}\,\sqrt{1-\frac{4m_e^2}{m_a^2}}\ll 1\,.
    \end{split}
\end{equation}
Therefore, the loop-induced decay of ALPs into electron-positron pairs is not relevant for our analysis and will be neglected in the following.

For ALP masses $m_a>3\,m_\pi$ (but still low enough for chiral perturbation theory to be valid), the QCD-ALP couplings in Eq.~\eqref{eq:LaaboveQCD} allow for decays into three pions $a\to3\,\pi^0$ and $a\to \pi^+\,\pi^-\,\pi^0$. These processes occur with a decay rate~\cite{Bauer:2017ris}
\begin{equation}
    \begin{aligned}
        \Gamma_{a3\pi} &= \frac{m_a m_\pi^4 \, (\Delta c_{ud})^2}{6144 \, \pi^3 f_\pi^2 f_a^2} \Theta(m_a - 3 m_\pi)\\
        &\quad \times \left[ g_0\left(\frac{m_\pi^2}{m_a^2}\right) + g_1\left(\frac{m_\pi^2}{m_a^2}\right) \right] \, ,
    \end{aligned}
\end{equation}
where 
\begin{equation}
        \begin{aligned}
        g_n(r) &= \frac{2 \cdot 6^n}{(1-r)^2} \int_{4r}^{(1-\sqrt{r})^2} d z \, \sqrt{1-\frac{4r}{z}} (z-r)^{2n} \\
        &\quad \times \sqrt{1+z^2+r^2-2z-2r-2zr} \, 
    \end{aligned}
\end{equation}
and
\begin{equation}
    \Delta c_{ud} = c_u - c_d + c_g \frac{m_d - m_u}{m_d + m_u} \,.
\end{equation}
The resulting partial decay length is of the order
\begin{equation}
\begin{split}
    \lambda_{3\pi} &\simeq \mathcal{O}\left(10^{10\dots14}\right) \cm \left(\frac{p_a}{m_a}\right) \\
    &\quad \times \left(\frac{m_a}{500 \MeV}\right)^{-1}
    \left(\frac{g_{ap}}{10^{-10}}\right)^{-2} \, ,
\end{split}
\end{equation}
where the exact values depend mostly on the coupling parameters in $\Delta c_{ud}$ but also on the value of the $g_i(m_\pi^2/m_a^2)$ (see Fig.~9 of Ref.~\cite{Bauer:2020jbp} for plots of these functions). Such values are comparable to the radii of SN progenitors.

Therefore, in this range of masses, the total ALP lifetime in its rest frame is given by
\begin{equation}
    \tau_a = \left(\Gamma_{a\gamma\gamma} + \Gamma_{a3\pi}\right)^{-1}\,.
\end{equation}
Depending on the values of the decay length ${\lambda_a= p_a/m_a\,\tau_a}$, ALPs can decay inside or outside the photosphere of the progenitor star with radius $R_{\rm env}$, giving rise to different signatures.
ALPs with $\lambda_a<R_{\mathrm{env}}$ will predominantly decay inside the SN progenitor star, depositing energy there, while those with $\lambda_a>R_{\mathrm{env}}$ mostly decay outside the volume of the star, leading to a potentially observable gamma-ray signal. We will discuss both of these scenarios in the following sections.

\subsection{Energy deposition in the SN envelope}
\label{sec:EnergyDeposition}
ALPs with masses $m_a\sim\mathcal{O}(10)-\mathcal{O}(100)\,\MeV$ can decay inside the SN mantle dumping a large amount of energy inside the volume of the progenitor star~\cite{Caputo:2022mah}. In particular, if ALP decays occur at radii $R$ between the PNS radius $R_{\mathrm{PNS}}$ and the envelope radius, the energy deposited by ALP decays could power the ejection of the outer layers of the mantle during the SN explosion event. Nevertheless, this energy deposition must not be larger than the predicted SN explosion energy, otherwise it would gravitationally unbind most of the progenitor mass, independently of neutrino heating or any other hypothetical explosion mechanism~\cite{Falk:1978kf,Sung:2019xie}. This argument provides a ``calorimetric'' constraint to ALP decays into photons. Moreover, as argued in Ref.~\cite{Caputo:2022mah}, to severely constrain such a scenario, it is helpful to employ a SN population with particularly low explosion energies as the most sensitive calorimeters~\cite{Stockinger:2020hse,Yang:2015ooa}. In this case, their low explosion energy requires that the energy released in the mantle by ALP radiative decays $E_{\rm dep}$ has not to exceed about $0.1$~B, where $1$ B (Bethe) $=10^{51}$~erg. According to Ref.~\cite{Caputo:2022mah}, the energy deposited in the SN envelope can be obtained as
\begin{equation}
    \begin{split}
        E_{\rm dep} = &4\pi \int dt \int_0^{R_{\rm PNS}} dr \, r^2 \int_{m_a}^\infty d\omega_a \, \omega_a \,\, \frac{d^2 n_a}{d {\omega_a}\,dt}(r,t,\omega_a^{\rm loc})\\  
        &\quad \times\left[ \exp\left(-\frac{R_{\rm PNS} -r}{\lambda_a}\right) - \exp\left(-\frac{R_{\rm env}-r}{\lambda_a}\right)\right]\,,
    \end{split}
    \label{eq:Emantle}
\end{equation}
where $d^2n_ a / d {\omega_a}\,dt$ is the spectral rate of change per unit volume of ALPs produced in nuclear processes, $\omega_a$ is the local production energy and the radial integration refers to the ALP production region. In this expression $\omega_a = \alpha \times \omega_a^{\rm loc}$ is the ALP energy measured by a distant observer, where $\alpha(r,t)$ is the lapse factor describing red-shifting of the local ALP energy, as well as the time dilation between the two references frames~\cite{Rampp:2002bq}.
The exponential terms select ALP decays occurring in the region between the PNS radius $R_{\rm PNS}\simeq30\,\km$ and the SN photosphere radius $R_{\rm env}$. We employ the value $R_{\rm env} = 5\times10^{13} \cm$, which is the typical radius of Red Supergiants at the end of their lives, which are the most common SN progenitors~\cite{Ouchi:2017cza,Goldberg:2020czt}.

Equation~\eqref{eq:Emantle} does not take into account that ALPs with strong couplings cannot only decay but could also be reabsorped by scattering events in the plasma before leaving the PNS. However, we note that this effect is expected to be relatively small in the part of the parameter space that we are studying here (especially since $g_{ap} < 10^{-8}$).
Still, the energy-deposition constraint is only approximate for the largest couplings considered here -- as is anyway the case since the energy transfer of relatively strongly coupled ALPs would presumably have an important impact on the explosion dynamics \cite{Caputo:2022mah}.

\begin{figure*}
    \centering
    \includegraphics[scale=0.6]{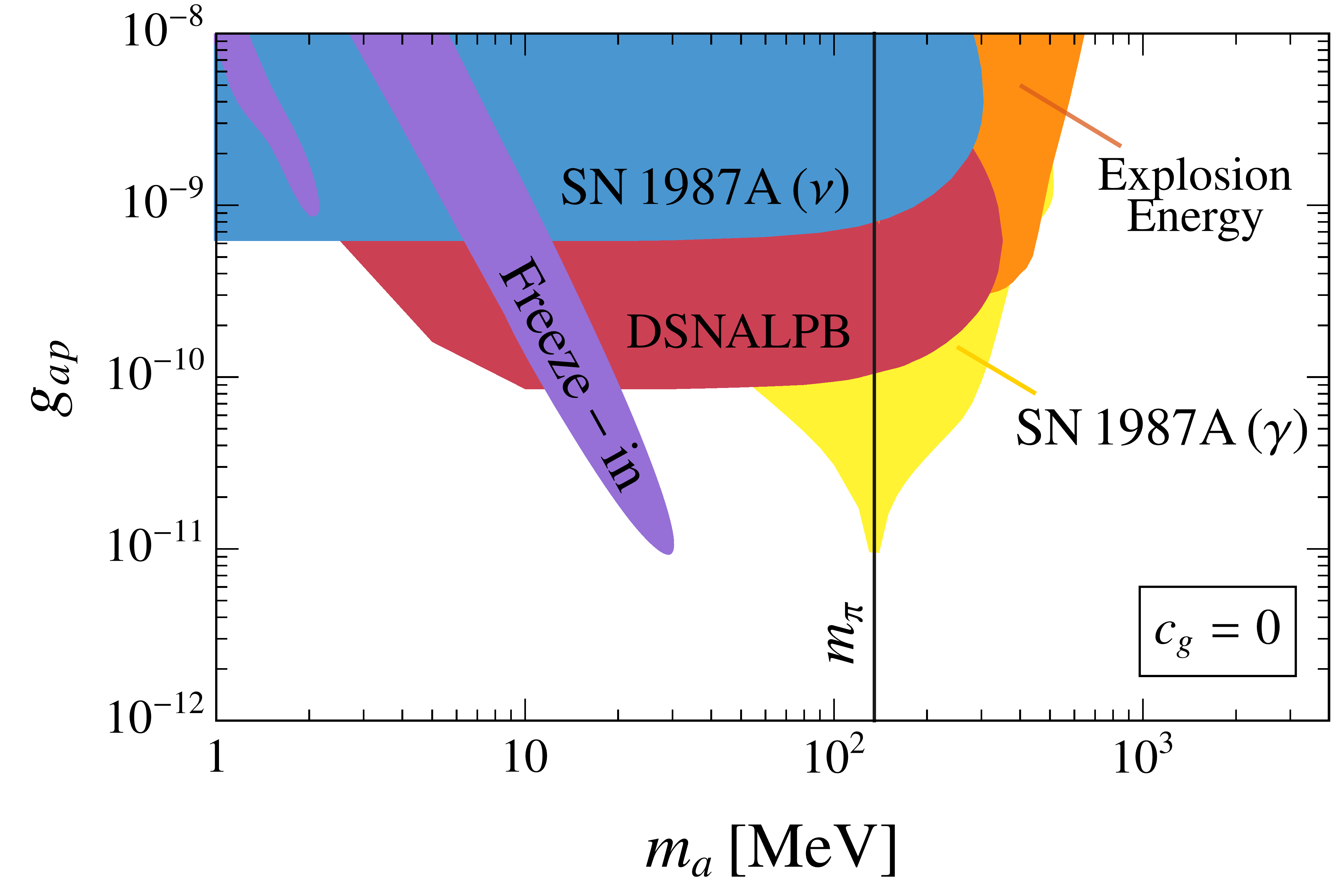}
    \centering
    \caption{Summary plot of the bounds 
    in the $c_g=0$ scenario. The blue region displays the SN 1987A cooling bound placed in Ref.~\cite{Lella:2023bfb}, while the violet region has been obtained by converting the limit on ALP-photon interactions placed in Ref.~\cite{Langhoff:2022bij} by searching for signatures of an irreducible ALP background to a constraint on $g_{ap}$. The other bounds have been calculated for this work, taking into account the induced photon coupling for ALPs coupled to nucleons.
    The red region is excluded by the non observation of any signature of a possible DSNALPB, the yellow contour depicts the range of parameters excluded by $\gamma$-observations in coincidence with SN 1987A, while the orange region is ruled out by requiring that ALPs do not deposit energy in the SN mantle in excess of observations.}
    \label{fig:bounds}
\end{figure*}

\begin{figure*}
    \centering
    \includegraphics[scale=0.6]{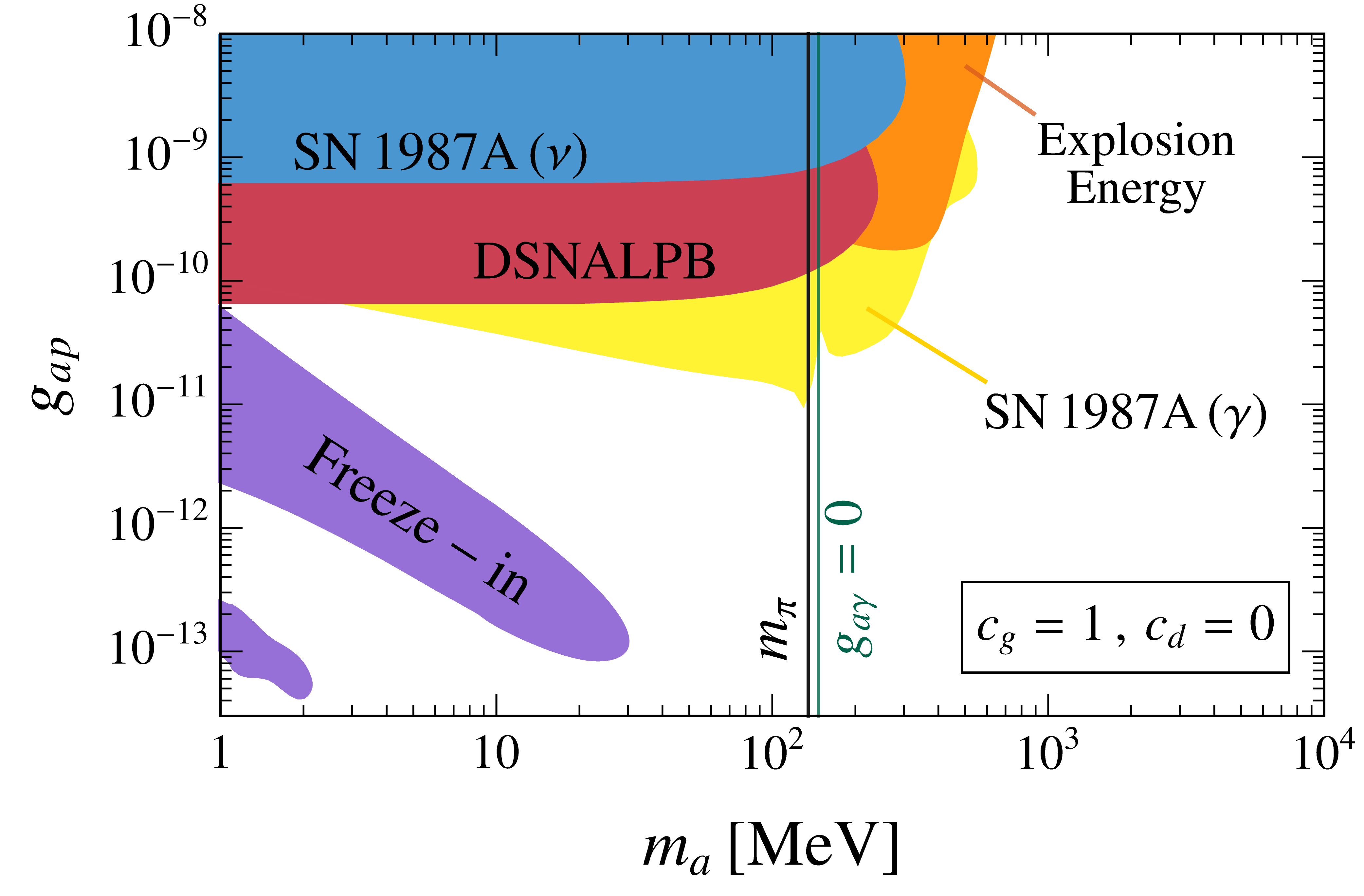}
    \centering
    \includegraphics[scale=0.6]{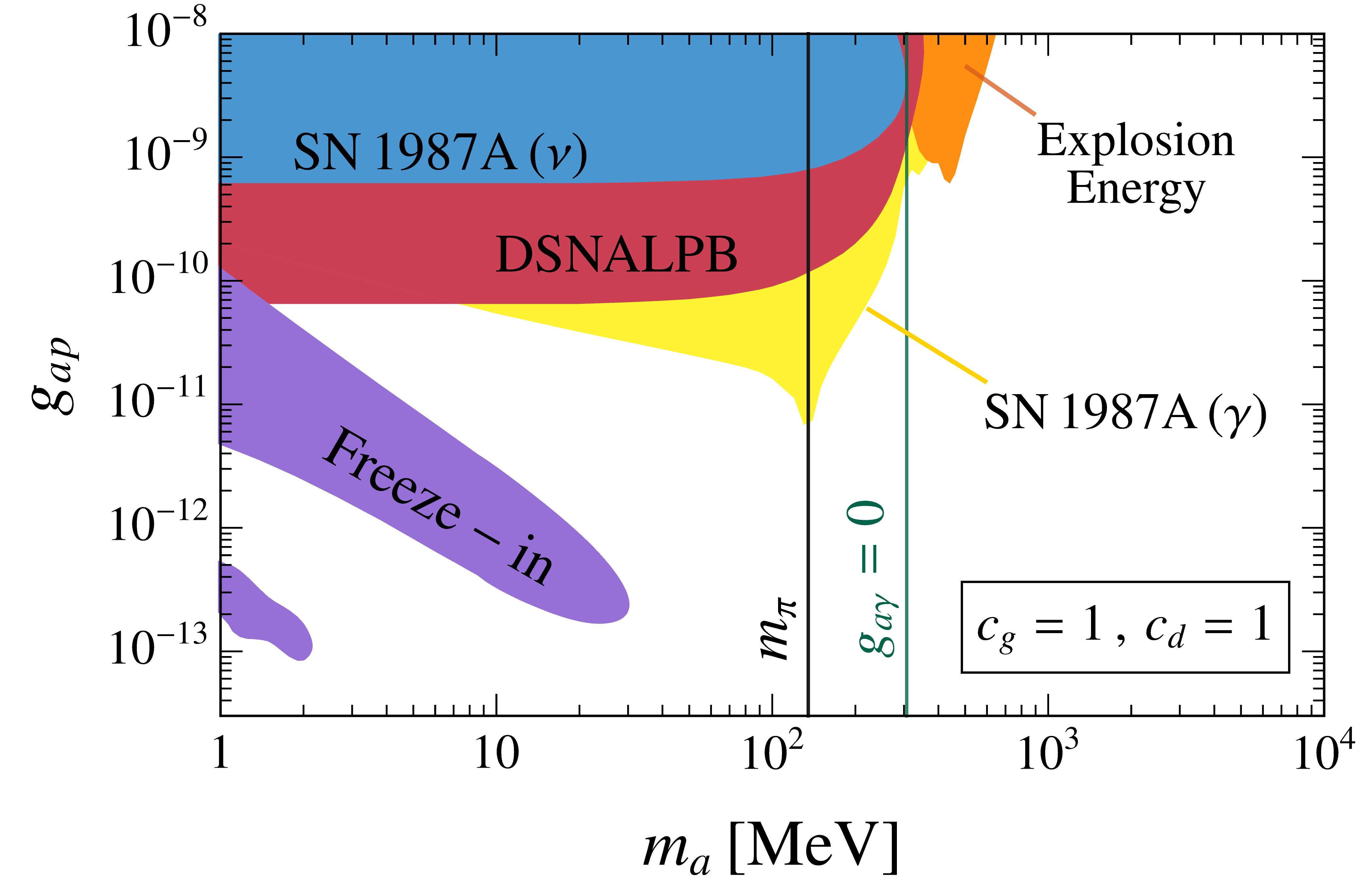}
    \caption{Summary plot of the bounds placed in this work in the case $c_g=1$. The color scheme for the contours is the same as in Fig.~\ref{fig:bounds}.
    The upper and lower panels refer to the $c_d=0$ and $c_d=1$ scenario, respectively.
    Here, the dark green vertical lines depict the values of the ALP mass for which the ALP-photon coupling vanishes in the considered cases. [Note that the very narrow kink in the DSNALPB constraint caused by the vanishing photon coupling is not visible due to the finite resolution of the plot.]}
    \label{fig:bounds2}
\end{figure*}

We highlight that our calculation of $E_{\rm dep}$ as shown in Eq.~\eqref{eq:Emantle} takes into account all relevant ALP decay processes, including $3\,\pi$ decays. Indeed, once produced, the $\pi^0$ decays almost immediately in photon pairs. On the other hand, charged pions would decay into (anti)muons with the corresponding (anti)neutrinos. However, in the progenitor star's mantle, the nuclear density is still high, so strong interactions with nuclei 
lead to the absorption of the charged pions in the nuclear medium before they can decay. Therefore, independently of the decay channel, the energy carried by decaying ALPs is entirely released in the mantle.
By requiring that the energy deposited inside the SN mantle is less than $0.1$~B, we excluded the orange regions of Figs.~\ref{fig:bounds} and~\ref{fig:bounds2} inside the ALP parameter space.


\subsection{ALP induced gamma-ray burst from SN 1987A}
\label{sec:decayBound}
If the ALPs produced in the core of a nearby SN can escape the photosphere of the progenitor star and decay at larger radii into gamma-rays, some of these would be able to reach detectors at Earth (see also Ref.~\cite{Muller:2023pip}).
The most constraining system to date is
SN 1987A, located in the Large Magellanic Cloud at a distance of $ d_{\rm SN} = 51.4 $~kpc. The Gamma-Ray Spectrometer on board the Solar Maximum Mission (SMM) satellite was taking data for $ \Delta t = 223 $~s after the first signal of SN 1987A, namely the neutrinos, reached Earth; no excess over the gamma-ray background was found \cite{Chupp:1989kx,Oberauer:1993yr}, and hence the existence of ALPs with certain parameters can be excluded \cite{Jaeckel:2017tud,Hoof:2022xbe,Muller:2023vjm}.

For SN 1987A as observed by SMM, we can use the simplified formula for the observable photon fluence \cite{Oberauer:1993yr,Jaffe:1995sw,Muller:2023vjm}:
\begin{equation} \label{eq:fluence}
\begin{aligned}
    F_\gamma = &\int_{m_a}^{\infty} d \omega_a \int_{\omega_\gamma^{\rm min}}^{\omega_\gamma^{\rm max}} d\omega_\gamma \, \Theta(\Delta\omega_\gamma) \frac{{\rm BR}_{a \to \gamma\gamma}}{2\pi d_{\rm SN}^2} \, p_a^{-1} \, \frac{dN_a}{d\omega_a}\\
    &\quad \times \left[ \exp\left(-\frac{m_a R_*}{\tau_a p_a}\right) - \exp\left(-\frac{2 \omega_\gamma \, \Delta t}{\tau_a m_a}\right) \right] \,,
\end{aligned}
\end{equation}
where $ R_* = 3 \times 10^{12} $~cm is the radius of the progenitor of SN 1987A (which, as a blue supergiant, was relatively small), $ d_{\rm SN} = 51.4 $~kpc is its distance to earth, ${d N_a}/{d\omega_a}$ is the total spectrum of ALPs as seen by a distant observer and the integration region is given by \cite{Muller:2023vjm}
\begin{equation}
\begin{aligned}
    \omega_\gamma^{{\rm min}}(p_a) &= \max \left( 25 \text{ MeV}, \, \frac{1}{2}(\omega_a - p_a), \, \frac{m_a^2 \, R_*}{2 \, p_a \, \Delta t} \right) \, , \\
    \omega_\gamma^{{\rm max}}(p_a) &= \min \left( 100 \text{ MeV}, \, \frac{1}{2}(\omega_a + p_a) \right) \, , \\
    \Delta \omega_\gamma(p_a) &= \omega_\gamma^{{\rm max}}(p_a) - \omega_\gamma^{{\rm min}}(p_a) \, .
\end{aligned}
\end{equation}
The ALP spectrum as observed by a far-away observer, including general relativistic corrections due to the strong gravity near the SN core, can be calculated as:
\begin{equation} \label{eq:alpSpectrum}
    \frac{d N_a}{d \omega_a} = 4\pi \int d t \int d r \, r^2 \, \alpha(r,t)^{-1} \dfrac{d^2 n_a}{d t \, d\omega_a}(r,t, \omega_a^{\rm loc}
    ) \,.
\end{equation}

Since no significant excess in the gamma-ray fluence has been observed by the SMM,  the fluence in Eq.~\eqref{eq:fluence} should not exceed a 3-$\sigma$ variation of the observed background data, i.e. $ F_\gamma < 1.78 \text{ cm}^{-2} $ \cite{Jaeckel:2017tud,Hoof:2022xbe}; the resulting bound is shown yellow in Figs.~\ref{fig:bounds}and~\ref{fig:bounds2}.

It was pointed out in Ref.~\cite{Diamond:2023scc} that in a small part of the parameter space seemingly excluded by the decay bound, a dense ``fireball'' QED-plasma could form through spatially and temporally concentrated ALP decays into photons. In this case, there would be no gamma-rays reaching Earth and the constraints from SMM observations would not apply. However, as pointed out in Ref.~\cite{Diamond:2023scc}, the respective parameter region would still typically be excluded by the non-observation of the fireballs by the Pierre Venus Orbiter. Hence, we do not expect that this argument would enlarge the parameter space allowed by observations.

\subsection{Diffuse SN ALP background}
\label{sec:DSNALPB}
ALPs produced in all past SN explosions in the observable Universe may have lead to a diffuse SN ALP background (DSNALPB)~\cite{Raffelt:2011ft}, analogous to the diffuse neutrino background~\cite{Beacom:2010kk}. This phenomenon and its observable consequences have been analyzed in Refs.~\cite{Eckner:2021kjb,Calore:2021hhn,Lella:2022uwi}, taking into account the production of ALPs with masses $m_a\sim\mathcal{O}(10)\,\MeV$ by means of $NN$ bremsstrahlung and pion conversions. Radiative decays of these heavy ALPs may have produced a contribution to the cosmic photon background, which is measured by gamma-ray telescopes such as \textit{Fermi}-LAT, and hence allows us to constrain the ALP parameters~\cite{Eckner:2021kjb}. Differently from the cited previous works, in our study, we analyze the scenario in which both ALP emission and decays are set by the ALP-nucleon coupling $g_{aN}$ and the induced, `irreducible' photon coupling as in Eq.~\eqref{eq:Cgammacg0} or Eq.~\eqref{eq:Cgammacg1}.

To obtain the total diffuse gamma-ray flux due to ALP decays in the DSNALPB we integrate over the red-shift $z$, and then sum up the contribution from all past core-collapse SNe~\cite{Calore:2020tjw}
\begin{equation}
    \frac{d\phi_\gamma^{\rm dif}}{dE_\gamma}=\int_0^\infty dz \, \left|\frac{dt}{dz}\right| (1+z) R_{\rm SN}(z) \frac{dN_\gamma(E_\gamma(1+z))}{dE_\gamma}\,,
\label{eq:GammaFluxDSNALPB}
\end{equation}
where $E_\gamma$ is the energy of the emitted photon. Here $R_{SN}(z)$ is the SN explosion rate, taken from Ref.~\cite{Priya:2017bmm}, with a total normalization for the core-collapse rate $R_{\rm cc}=1.25\times10^{-4}\text{yr}^{-1}\Mpc^{-3}$ and computed assuming an average progenitor mass of about $18$~$M_{\odot}$, as in the SN simulation employed in this work. Furthermore, $|dt/dz|^{-1}$ is given by $|dt/dz|^{-1}=H_0(1+z)[\Omega_\Lambda+\Omega_M(1+z)^3]^{\frac{1}{2}}$ with the cosmological parameters $H_0=67.4 \km \s^{-1} \Mpc^{-1}$, $\Omega_\Lambda=0.7$ and $\Omega_M=0.3$.
Taking into account that the daughter photons can have any energy in the range of $(\omega_a - p_a)/2 < E_\gamma < (\omega_a + p_a)/2$, and that each energy corresponds to a unique angle between photon and original ALP momentum, the isotropic gamma-ray flux induced by ALP decays from a single SN at redshift $z$ is
\begin{equation}
\begin{aligned}
    \frac{d N_\gamma(E_\gamma)}{d E_\gamma}&=\int_{p_a^{\rm min}}^{\infty} \frac{d p_a}{\omega_a} \, 2 \times {\rm BR}_{a \to \gamma\gamma}\,\,\frac{d N_a}{d \omega_a}\\
    &\quad \times \left[ \exp\left(-\frac{R'_{\rm env} m_a}{p_a \tau_a}\right) - \exp\left(-\frac{d(z) m_a}{p_a \tau_a}\right) \right]\,,
\end{aligned}
\end{equation}
where $d(z)$ is the cosmological distance of a SN that occurred at redshift $z$, $ p_a^{\rm min} = E_\gamma + {m_a^2}/{4 E_\gamma} $ is the minimal ALP momentum contributing to the flux at photon energy $E_\gamma$ and the ALP spectrum $d N_a / d \omega_a$ is given in Eq.~\eqref{eq:alpSpectrum}. We highlight that here we use $R'_{\rm env} = 10^{14}\,\cm$ to assure that ALP decays always occur outside from all the considered SNe, which might admit progenitors with radii even larger than $R_{\rm env} = 5\times10^{13}\,\cm$. This assumption leads to more conservative constraints.

Following Ref.~\cite{Calore:2020tjw}, to constrain this scenario we have employed the isotropic gamma-ray background measurements provided by the \textit{Fermi}-LAT Collaboration, by means of  Pass 8 R3 processed data (8-yr dataset) for the ULTRACLEANVETO event class section. In the range of energies $E_\gamma\gtrsim50 \MeV$, the \textit{Fermi}-LAT 
data for the diffuse gamma-ray flux can be fitted as
\begin{equation}
    \frac{d\phi_\gamma(E_\gamma)}{dE_\gamma}\simeq2.2\times10^{-3}\left(\frac{E_\gamma}{\MeV}\right)^{-2.2} \MeV^{-1} \cm^{-2} \s^{-1} \text{sr}^{-1}\,.
\label{eq:FermiData}
\end{equation}
On the other hand, fluxes at lower energies have to be compared to the measurements from the COMPTEL experiment~\cite{2008ICRC....3.1085S}
\begin{equation}
    \frac{d\phi_\gamma(E_\gamma)}{dE_\gamma}\simeq1.05\times10^{-4}\left(\frac{E_\gamma}{5\,\MeV}\right)^{-2.4} \MeV^{-1} \cm^{-2} \s^{-1} \text{sr}^{-1}\,.
\label{eq:CompetelData}
\end{equation}
The contribution from the DSNALPB in Eq.~\eqref{eq:GammaFluxDSNALPB} cannot be larger than the observed flux: $d\phi_\gamma^{\rm dif} / dE_\gamma < d\phi_\gamma / dE_\gamma$.
This argument allows us to constrain the red areas of the parameter spaces shown in Figs.~\ref{fig:bounds} and \ref{fig:bounds2}.

\section{Discussion and Results}
\label{sec:results}

\subsection{$c_g=0$}
The first case we want to analyze is $c_g=0$, corresponding to the scenario of a vanishing ALP-gluon coupling in Eq.~\eqref{eq:LaaboveQCD}. Under this assumption, the induced ALP-photon coupling shows a strong dependence on the ALP mass~[see Eq.~\eqref{eq:photon_coupling2}]. In particular, for masses $m_a\ll m_\pi$, the photon coupling is suppressed by factors $\mathcal{O}(m_a^2/m_\pi^2)$ leading to inefficient ALP-photon interactions. This effect has a significant impact on the relevant ALP phenomenology, since the ALP decay rate is strongly reduced in this range of masses. Moreover, an additional suppression due to the mass dependence of the ALP decay rate $\Gamma_{a\gamma\gamma} \propto m_a^3$ has to be considered. As a consequence, all the related bounds tend to relax in the mass range $m_a\lesssim10\,\MeV$ since too long-lived ALPs will decay too late to deposit energy into the stellar mantle, to source an observable gamma-ray signal on their way to Earth from SN 1987A, or to decay in the observable Universe.

Fig.~\ref{fig:bounds} summarizes all the constraints introduced in this work in the case $c_g=0$ by employing the arguments discussed in the previous sections. The black line highlights the presence of the pole in Eqs.~\eqref{eq:photon_coupling2} in coincidence with the pion mass. As discussed in Sec.~\ref{sec:ALPinteractions}, the pole itself is unphysical but Eq.~\eqref{eq:photon_coupling2} only breaks down in a very narrow region around $m_\pi$. Violet contours display constraints from the irreducible cosmic ALP background established through freeze-in as described in Ref.~\cite{Langhoff:2022bij}. The two regions are excluded because of decays of the ALP background on cosmological time scales which might induce distortions in the CMB spectrum or observable diffuse X-ray signatures in XMM-Newton. Here the constraints on the photon coupling introduced in Ref.~\cite{Langhoff:2022bij} are converted to constraints on $g_{ap}$ by virtue of Eqs.~\eqref{eq:photon_coupling2}, which assumes that the effective photon coupling is induced by the underlying QCD coupling.

The red area depicts the region of the parameter space excluded by the DSNALPB argument illustrated in Sec.~\ref{sec:DSNALPB}. In particular, it can be employed to exclude ALP-proton couplings $g_{ap}\gtrsim8.5\times10^{-11}$ for $10\lesssim m_a \lesssim100\,\MeV$, extending the exclusion region from the SN-cooling argument~\cite{Lella:2023bfb,Lella:2022uwi} by one order of magnitude. As discussed above, the bound relaxes for $m_a\lesssim10\,\MeV$ as a consequence of the mass dependence of $g_{a\gamma}$ and the decay rate. On the other hand, at $m_a\gtrsim200\,\MeV$ smaller regions of the parameter space are excluded, since the fraction of ALPs decaying inside the SN envelope becomes larger and their production in the PNS is Boltzmann suppressed. 
The parameter space for ALPs with masses $m_a\gtrsim200\MeV$ can be constrained by looking at the energy deposited in the SN volume by decaying ALPs, as described in Sec.~\ref{sec:EnergyDeposition}. The orange region in the Fig.~\ref{fig:bounds} shows that this argument may rule out ALP-proton couplings $g_{ap}\gtrsim2.5\times10^{-10}$ for $200\lesssim m_a \lesssim300\,\MeV$, enlarging the mass range probed by the SN cooling argument.
In the yellow area of the parameter space in Fig.~\ref{fig:bounds}, the existence of ALPs would have led to a gamma-ray signal following SN 1987A observable by SMM as described in Sec.~\ref{sec:decayBound}. Together with the cosmological freeze-in constraint, this bound reaches the smallest couplings -- nearly two orders of magnitude lower than the cooling argument. However, due to the strong mass dependence of the effective photon coupling in the $c_g = 0$ case, only relatively heavy ALPs in the range $ 60 \MeV \lesssim m_a \lesssim 400 \MeV$ are excluded by this bounds only.

\subsection{$c_g=1$}
Fig.~\ref{fig:bounds2} summarizes the constraints introduced in this work in the $c_g=1$ scenario, where the upper and lower panel refer to the benchmark cases $c_d=0$ and $c_d=1$, respectively.
The effective, irreducible photon coupling in the case $c_g=1$ does not suffer from the ALP-mass suppression for $m_a\lesssim10\,\MeV$ since the ALP-gluon coupling always induces a sizable mass-independent contribution~[see Eq.~\eqref{eq:Cgammacg1}]. Nevertheless, in Sec.~\ref{sec:ALPinteractions} we have discussed that the ALP-photon coupling could actually vanish for some peculiar values of the ALP mass $m_a \simeq147\,\MeV$ and $m_a \simeq310\,\MeV$, which are depicted as a green lines in Fig.~\ref{fig:bounds2} for both the $c_d=0$ and $c_d=1$ cases. At these values of the mass, ALPs coupled to protons and gluons do not show any induced coupling to photons and all the constraints related to ALP-photon interactions are relaxed. However, except for ALP masses fine-tuned at the level of $\mathcal{O}(1)\,\MeV$ to these values, the ALP-photon coupling is sizable.

Differently to the $c_g=0$ scenario, the bound associated with a possible DSNALPB from all the past SNe is mostly flat in this case in the range of masses $m_a\gtrsim1\,\MeV$, excluding ALP couplings $g_{ap}\gtrsim6.5\times10^{-11}$. Indeed, even at low masses, the ALP-nucleon coupling induces ALP decays efficiently enough to produce a diffuse flux of photons, which saturates the condition described in Sec.~\ref{sec:DSNALPB}. Furthermore, in the $c_d=1$ case, we can probe larger values of the mass than in the $c_d=0$ scenario. Indeed, Fig.~\ref{fig:Cgamma} suggests that the induced ALP-photon coupling is suppressed for $250\lesssim m_a\lesssim 320\,\MeV$ in the $c_d=1$ case. Thus, in this mass range, ALP-photon interactions are inefficient enough for the radiative decays to happen outside the SN envelope, while in the $c_d=0$ case, most ALPs decay in the envelope and hence do not contribute to the DSNALPB.
This effect can also be seen in the explosion energy bound. The relaxation of the constraint in the $c_d=1$ case with respect to $c_d=0$ is determined by larger values of the decay lengths in the range of masses leading to inefficient ALP-photon interactions. As a consequence, most of the ALPs decay outside the SN mantle, and a smaller fraction could take part in the energy deposition phenomenon. Therefore, the explosion energy argument can rule out $g_{ap}\gtrsim1.5\times10^{-10}$ for $200\lesssim m_a \lesssim300\,\MeV$ in the $c_d=0$ case and only $g_{ap}\gtrsim10^{-9}$ for $300\lesssim m_a \lesssim400\,\MeV$ in the $c_d=1$ scenario.
Finally, the gamma-ray constraint from SN 1987A can, in this case, cover parameter space over a wide range of masses since even for small masses, more ALPs decay between the SN and Earth as compared to the case of a smaller $g_{a\gamma}$ for $c_g = 0$.
With all constraints derived in this work taken together, the cooling argument is fully superseded in the full parameter space that we study here.

\section{Conclusions}
\label{sec:conclusions}

Our investigation of QCD ALPs has shed light on the interplay between ALP properties at high-energy scales and the associated low-energy phenomenology, carefully taking into account the full set of interactions induced by fundamental couplings. In this spirit, we have explored the origin of ALP-nucleon interactions, clarifying how these interaction vertices appear as the manifestation of an effective field theory at higher scales with 
ALP couplings to quarks and gluons. 
We pointed out that an `irreducible' ALP-photon coupling naturally emerges in this scenario, and it must be taken into account in astrophysical searches.

The presence of both of these couplings opens several possibilities for phenomenology, especially for MeV-scale ALPs produced in SNe.
Our analysis, building upon recent developments in the study of ALP production during SN cooling phases~\cite{Lella:2023bfb}, which proceeds very efficiently via the ALP-nucleon coupling, also considers the role of the induced ALP-photon coupling arising in these models, leading to ALP decays into photon pairs. This decay channel could induce possible signatures in 
astrophysical observables related to SN events, such as additional explosion energy deposited in the SN mantle, an ALP-induced gamma-ray burst, or a contribution to the DSNALPB.
Using these arguments, in this work, we have set various constraints on the ALP-proton coupling, down to $g_{ap}\sim 10^{-10}-10^{-11}$ in the mass range ${1~\MeV\lesssim m_{a}\lesssim500~\MeV}$, ruling out regions of the parameter space associated to ALPs coupled to nuclear matter that have never been probed before.

\section*{Acknowledgements}
This work was initiated during short-term scientific mission of A.~L.'s at Stockholm University, funded by COST Action COSMIC WISPers CA21106.
We warmly thank Thomas Janka for giving us access to the {\tt GARCHING} group archive, and Luca di Luzio for useful discussions on the topic.
This article/publication is based upon work from COST Action COSMIC WISPers CA21106, supported by COST (European Cooperation in Science and Technology).
This work is (partially) supported
by ICSC – Centro Nazionale di Ricerca in High Performance Computing,
Big Data and Quantum Computing, funded by European Union - NextGenerationEU.
The work of D.~M., E.~R.~and P.~C. is supported by the Swedish Research Council (VR) under Grants No. 2018-03641 and No. 2019-02337.
The work of A.~L. was partially supported by the Research Grant No. 2022E2J4RK "PANTHEON: Perspectives in Astroparticle and
Neutrino THEory with Old and New messengers" under the program PRIN 2022 funded by the Italian Ministero dell’Universit\`a e della Ricerca (MUR).

\bibliographystyle{bibi.bst}
\bibliography{references.bib}

\providecommand{\href}[2]{#2}\begingroup\raggedright\begin{thebibliography}{100}

\bibitem{Jaeckel:2010ni}
J.~Jaeckel and A.~Ringwald, \emph{{The Low-Energy Frontier of Particle Physics}}, \href{https://doi.org/10.1146/annurev.nucl.012809.104433}{\emph{Ann. Rev. Nucl. Part. Sci.} {\bfseries 60} (2010) 405} [\href{https://arxiv.org/abs/1002.0329}{{\ttfamily 1002.0329}}].

\bibitem{DiLuzio:2020wdo}
L.~Di~Luzio, M.~Giannotti, E.~Nardi and L.~Visinelli, \emph{{The landscape of QCD axion models}}, \href{https://doi.org/10.1016/j.physrep.2020.06.002}{\emph{Phys. Rept.} {\bfseries 870} (2020) 1} [\href{https://arxiv.org/abs/2003.01100}{{\ttfamily 2003.01100}}].

\bibitem{Weinberg:1977ma}
S.~Weinberg, \emph{{A New Light Boson?}}, \href{https://doi.org/10.1103/PhysRevLett.40.223}{\emph{Phys. Rev. Lett.} {\bfseries 40} (1978) 223}.

\bibitem{Wilczek:1977pj}
F.~Wilczek, \emph{{Problem of Strong $P$ and $T$ Invariance in the Presence of Instantons}}, \href{https://doi.org/10.1103/PhysRevLett.40.279}{\emph{Phys. Rev. Lett.} {\bfseries 40} (1978) 279}.

\bibitem{Peccei:1977hh}
R.~D. Peccei and H.~R. Quinn, \emph{{CP Conservation in the Presence of Instantons}}, \href{https://doi.org/10.1103/PhysRevLett.38.1440}{\emph{Phys. Rev. Lett.} {\bfseries 38} (1977) 1440}.

\bibitem{Peccei:1977ur}
R.~D. Peccei and H.~R. Quinn, \emph{{Constraints Imposed by CP Conservation in the Presence of Instantons}}, \href{https://doi.org/10.1103/PhysRevD.16.1791}{\emph{Phys. Rev. D} {\bfseries 16} (1977) 1791}.

\bibitem{Svrcek:2006yi}
P.~Svrcek and E.~Witten, \emph{{Axions In String Theory}}, \href{https://doi.org/10.1088/1126-6708/2006/06/051}{\emph{JHEP} {\bfseries 06} (2006) 051} [\href{https://arxiv.org/abs/hep-th/0605206}{{\ttfamily hep-th/0605206}}].

\bibitem{Cicoli:2012sz}
M.~Cicoli, M.~Goodsell and A.~Ringwald, \emph{{The type IIB string axiverse and its low-energy phenomenology}}, \href{https://doi.org/10.1007/JHEP10(2012)146}{\emph{JHEP} {\bfseries 10} (2012) 146} [\href{https://arxiv.org/abs/1206.0819}{{\ttfamily 1206.0819}}].

\bibitem{Halverson:2019kna}
J.~Halverson, C.~Long, B.~Nelson and G.~Salinas, \emph{{Axion reheating in the string landscape}}, \href{https://doi.org/10.1103/PhysRevD.99.086014}{\emph{Phys. Rev. D} {\bfseries 99} (2019) 086014} [\href{https://arxiv.org/abs/1903.04495}{{\ttfamily 1903.04495}}].

\bibitem{Preskill:1982cy}
J.~Preskill, M.~B. Wise and F.~Wilczek, \emph{{Cosmology of the Invisible Axion}}, \href{https://doi.org/10.1016/0370-2693(83)90637-8}{\emph{Phys. Lett. B} {\bfseries 120} (1983) 127}.

\bibitem{Abbott:1982af}
L.~F. Abbott and P.~Sikivie, \emph{{A Cosmological Bound on the Invisible Axion}}, \href{https://doi.org/10.1016/0370-2693(83)90638-X}{\emph{Phys. Lett. B} {\bfseries 120} (1983) 133}.

\bibitem{Dine:1982ah}
M.~Dine and W.~Fischler, \emph{{The Not So Harmless Axion}}, \href{https://doi.org/10.1016/0370-2693(83)90639-1}{\emph{Phys. Lett. B} {\bfseries 120} (1983) 137}.

\bibitem{Lawson:2019brd}
M.~Lawson, A.~J. Millar, M.~Pancaldi, E.~Vitagliano and F.~Wilczek, \emph{{Tunable axion plasma haloscopes}}, \href{https://doi.org/10.1103/PhysRevLett.123.141802}{\emph{Phys. Rev. Lett.} {\bfseries 123} (2019) 141802} [\href{https://arxiv.org/abs/1904.11872}{{\ttfamily 1904.11872}}].

\bibitem{BREAD:2021tpx}
{\scshape BREAD} Collaboration, J.~Liu et~al., \emph{{Broadband Solenoidal Haloscope for Terahertz Axion Detection}}, \href{https://doi.org/10.1103/PhysRevLett.128.131801}{\emph{Phys. Rev. Lett.} {\bfseries 128} (2022) 131801} [\href{https://arxiv.org/abs/2111.12103}{{\ttfamily 2111.12103}}].

\bibitem{DMRadio:2022pkf}
{\scshape DMRadio} Collaboration, L.~Brouwer et~al., \emph{{Projected sensitivity of DMRadio-m3: A search for the QCD axion below 1\,\,\ensuremath{\mu}eV}}, \href{https://doi.org/10.1103/PhysRevD.106.103008}{\emph{Phys. Rev. D} {\bfseries 106} (2022) 103008} [\href{https://arxiv.org/abs/2204.13781}{{\ttfamily 2204.13781}}].

\bibitem{Raffelt:1987im}
G.~Raffelt and L.~Stodolsky, \emph{{Mixing of the Photon with Low Mass Particles}}, \href{https://doi.org/10.1103/PhysRevD.37.1237}{\emph{Phys. Rev. D} {\bfseries 37} (1988) 1237}.

\bibitem{Irastorza:2018dyq}
I.~G. Irastorza and J.~Redondo, \emph{{New experimental approaches in the search for axion-like particles}}, \href{https://doi.org/10.1016/j.ppnp.2018.05.003}{\emph{Prog. Part. Nucl. Phys.} {\bfseries 102} (2018) 89} [\href{https://arxiv.org/abs/1801.08127}{{\ttfamily 1801.08127}}].

\bibitem{Sikivie:2020zpn}
P.~Sikivie, \emph{{Invisible Axion Search Methods}}, \href{https://doi.org/10.1103/RevModPhys.93.015004}{\emph{Rev. Mod. Phys.} {\bfseries 93} (2021) 015004} [\href{https://arxiv.org/abs/2003.02206}{{\ttfamily 2003.02206}}].

\bibitem{Raffelt:1985nk}
G.~G. Raffelt, \emph{{ASTROPHYSICAL AXION BOUNDS DIMINISHED BY SCREENING EFFECTS}}, \href{https://doi.org/10.1103/PhysRevD.33.897}{\emph{Phys. Rev. D} {\bfseries 33} (1986) 897}.

\bibitem{Caputo:2024oqc}
A.~Caputo and G.~Raffelt, \emph{{Astrophysical Axion Bounds: The 2024 Edition}}, \href{https://doi.org/10.22323/1.454.0041}{\emph{PoS} {\bfseries COSMICWISPers} (2024) 041} [\href{https://arxiv.org/abs/2401.13728}{{\ttfamily 2401.13728}}].

\bibitem{DiLuzio:2021ysg}
L.~Di~Luzio, M.~Fedele, M.~Giannotti, F.~Mescia and E.~Nardi, \emph{{Stellar evolution confronts axion models}}, \href{https://doi.org/10.1088/1475-7516/2022/02/035}{\emph{JCAP} {\bfseries 02} (2022) 035} [\href{https://arxiv.org/abs/2109.10368}{{\ttfamily 2109.10368}}].

\bibitem{Wouters:2013hua}
D.~Wouters and P.~Brun, \emph{{Constraints on Axion-like Particles from X-Ray Observations of the Hydra Galaxy Cluster}}, \href{https://doi.org/10.1088/0004-637X/772/1/44}{\emph{Astrophys. J.} {\bfseries 772} (2013) 44} [\href{https://arxiv.org/abs/1304.0989}{{\ttfamily 1304.0989}}].

\bibitem{Berg:2016ese}
M.~Berg, J.~P. Conlon, F.~Day, N.~Jennings, S.~Krippendorf, A.~J. Powell and M.~Rummel, \emph{{Constraints on Axion-Like Particles from X-ray Observations of NGC1275}}, \href{https://doi.org/10.3847/1538-4357/aa8b16}{\emph{Astrophys. J.} {\bfseries 847} (2017) 101} [\href{https://arxiv.org/abs/1605.01043}{{\ttfamily 1605.01043}}].

\bibitem{Reynolds:2019uqt}
C.~S. Reynolds, M.~C.~D. Marsh, H.~R. Russell, A.~C. Fabian, R.~Smith, F.~Tombesi and S.~Veilleux, \emph{{Astrophysical limits on very light axion-like particles from Chandra grating spectroscopy of NGC 1275}}, \href{https://doi.org/10.3847/1538-4357/ab6a0c}{\emph{Astrophys. J.} {\bfseries 890} (2020) 59} [\href{https://arxiv.org/abs/1907.05475}{{\ttfamily 1907.05475}}].

\bibitem{Reynes:2021bpe}
J.~S. Reyn\'es, J.~H. Matthews, C.~S. Reynolds, H.~R. Russell, R.~N. Smith and M.~C.~D. Marsh, \emph{{New constraints on light axion-like particles using Chandra transmission grating spectroscopy of the powerful cluster-hosted quasar H1821+643}}, \href{https://doi.org/10.1093/mnras/stab3464}{\emph{Mon. Not. Roy. Astron. Soc.} {\bfseries 510} (2021) 1264} [\href{https://arxiv.org/abs/2109.03261}{{\ttfamily 2109.03261}}].

\bibitem{Fermi-LAT:2016nkz}
{\scshape Fermi-LAT} Collaboration, M.~Ajello et~al., \emph{{Search for Spectral Irregularities due to Photon\textendash{}Axionlike-Particle Oscillations with the Fermi Large Area Telescope}}, \href{https://doi.org/10.1103/PhysRevLett.116.161101}{\emph{Phys. Rev. Lett.} {\bfseries 116} (2016) 161101} [\href{https://arxiv.org/abs/1603.06978}{{\ttfamily 1603.06978}}].

\bibitem{Davies:2022wvj}
J.~Davies, M.~Meyer and G.~Cotter, \emph{{Constraints on axionlike particles from a combined analysis of three flaring Fermi flat-spectrum radio quasars}}, \href{https://doi.org/10.1103/PhysRevD.107.083027}{\emph{Phys. Rev. D} {\bfseries 107} (2023) 083027} [\href{https://arxiv.org/abs/2211.03414}{{\ttfamily 2211.03414}}].

\bibitem{Li:2021gxs}
H.-J. Li, X.-J. Bi and P.-F. Yin, \emph{{Searching for axion-like particles with the blazar observations of MAGIC and Fermi-LAT *}}, \href{https://doi.org/10.1088/1674-1137/ac6d4f}{\emph{Chin. Phys. C} {\bfseries 46} (2022) 085105} [\href{https://arxiv.org/abs/2110.13636}{{\ttfamily 2110.13636}}].

\bibitem{Meyer:2016wrm}
M.~Meyer, M.~Giannotti, A.~Mirizzi, J.~Conrad and M.~A. S\'anchez-Conde, \emph{{Fermi Large Area Telescope as a Galactic Supernovae Axionscope}}, \href{https://doi.org/10.1103/PhysRevLett.118.011103}{\emph{Phys. Rev. Lett.} {\bfseries 118} (2017) 011103} [\href{https://arxiv.org/abs/1609.02350}{{\ttfamily 1609.02350}}].

\bibitem{Meyer:2020vzy}
M.~Meyer and T.~Petrushevska, \emph{{Search for Axionlike-Particle-Induced Prompt $\gamma$-Ray Emission from Extragalactic Core-Collapse Supernovae with the $Fermi$ Large Area Telescope}}, \href{https://doi.org/10.1103/PhysRevLett.124.231101}{\emph{Phys. Rev. Lett.} {\bfseries 124} (2020) 231101} [\href{https://arxiv.org/abs/2006.06722}{{\ttfamily 2006.06722}}]. [Erratum: Phys.Rev.Lett. 125, 119901 (2020)].

\bibitem{Jacobsen:2022swa}
S.~Jacobsen, T.~Linden and K.~Freese, \emph{{Constraining axion-like particles with HAWC observations of TeV blazars}}, \href{https://doi.org/10.1088/1475-7516/2023/10/009}{\emph{JCAP} {\bfseries 10} (2023) 009} [\href{https://arxiv.org/abs/2203.04332}{{\ttfamily 2203.04332}}].

\bibitem{MAGIC:2024arq}
{\scshape MAGIC, INFN MAGIC Group: INFN Sezione di Torino, Universit\`a degli Studi di Torino, I-10125 Torino, Italy, INFN MAGIC Group: INFN Sezione di Bari, Dipartimento Interateneo di Fisica dell'Universit\`a e del Politecnico di Bari, I-70125 Bari, Italy, Armenian MAGIC Group: ICRANet-Armenia, 0019 Yerevan, Armenia, INFN MAGIC Group: INFN Sezione di Perugia, I-06123 Perugia, Italy, INFN MAGIC Group: INFN Roma Tor Vergata, I-00133 Roma, Italy} Collaboration, H.~Abe et~al., \emph{{Constraints on axion-like particles with the Perseus Galaxy Cluster with MAGIC}}, \href{https://doi.org/10.1016/j.dark.2024.101425}{\emph{Phys. Dark Univ.} {\bfseries 44} (2024) 101425} [\href{https://arxiv.org/abs/2401.07798}{{\ttfamily 2401.07798}}].

\bibitem{Mastrototaro:2022kpt}
L.~Mastrototaro, P.~Carenza, M.~Chianese, D.~F.~G. Fiorillo, G.~Miele, A.~Mirizzi and D.~Montanino, \emph{{Constraining axion-like particles with the diffuse gamma-ray flux measured by the Large High Altitude Air Shower Observatory}}, \href{https://doi.org/10.1140/epjc/s10052-022-10979-6}{\emph{Eur. Phys. J. C} {\bfseries 82} (2022) 1012} [\href{https://arxiv.org/abs/2206.08945}{{\ttfamily 2206.08945}}].

\bibitem{Calore:2023srn}
F.~Calore, P.~Carenza, C.~Eckner, M.~Giannotti, G.~Lucente, A.~Mirizzi and F.~Sivo, \emph{{Uncovering axionlike particles in supernova gamma-ray spectra}}, \href{https://doi.org/10.1103/PhysRevD.109.043010}{\emph{Phys. Rev. D} {\bfseries 109} (2024) 043010} [\href{https://arxiv.org/abs/2306.03925}{{\ttfamily 2306.03925}}].

\bibitem{Burrows:1988ah}
A.~Burrows, M.~S. Turner and R.~P. Brinkmann, \emph{{Axions and SN 1987a}}, \href{https://doi.org/10.1103/PhysRevD.39.1020}{\emph{Phys. Rev. D} {\bfseries 39} (1989) 1020}.

\bibitem{Burrows:1990pk}
A.~Burrows, M.~T. Ressell and M.~S. Turner, \emph{{Axions and SN1987A: Axion trapping}}, \href{https://doi.org/10.1103/PhysRevD.42.3297}{\emph{Phys. Rev. D} {\bfseries 42} (1990) 3297}.

\bibitem{Carenza:2019pxu}
P.~Carenza, T.~Fischer, M.~Giannotti, G.~Guo, G.~Mart\'\i{}nez-Pinedo and A.~Mirizzi, \emph{{Improved axion emissivity from a supernova via nucleon-nucleon bremsstrahlung}}, \href{https://doi.org/10.1088/1475-7516/2019/10/016}{\emph{JCAP} {\bfseries 10} (2019) 016} [\href{https://arxiv.org/abs/1906.11844}{{\ttfamily 1906.11844}}]. [Erratum: JCAP 05, E01 (2020)].

\bibitem{Carenza:2021ebx}
P.~Carenza, M.~Lattanzi, A.~Mirizzi and F.~Forastieri, \emph{{Thermal axions with multi-eV masses are possible in low-reheating scenarios}}, \href{https://doi.org/10.1088/1475-7516/2021/07/031}{\emph{JCAP} {\bfseries 07} (2021) 031} [\href{https://arxiv.org/abs/2104.03982}{{\ttfamily 2104.03982}}].

\bibitem{Fischer:2021jfm}
T.~Fischer, P.~Carenza, B.~Fore, M.~Giannotti, A.~Mirizzi and S.~Reddy, \emph{{Observable signatures of enhanced axion emission from protoneutron stars}}, \href{https://doi.org/10.1103/PhysRevD.104.103012}{\emph{Phys. Rev. D} {\bfseries 104} (2021) 103012} [\href{https://arxiv.org/abs/2108.13726}{{\ttfamily 2108.13726}}].

\bibitem{Lella:2022uwi}
A.~Lella, P.~Carenza, G.~Lucente, M.~Giannotti and A.~Mirizzi, \emph{{Protoneutron stars as cosmic factories for massive axionlike particles}}, \href{https://doi.org/10.1103/PhysRevD.107.103017}{\emph{Phys. Rev. D} {\bfseries 107} (2023) 103017} [\href{https://arxiv.org/abs/2211.13760}{{\ttfamily 2211.13760}}].

\bibitem{Lella:2023bfb}
A.~Lella, P.~Carenza, G.~Co', G.~Lucente, M.~Giannotti, A.~Mirizzi and T.~Rauscher, \emph{{Getting the most on supernova axions}}, \href{https://doi.org/10.1103/PhysRevD.109.023001}{\emph{Phys. Rev. D} {\bfseries 109} (2024) 023001} [\href{https://arxiv.org/abs/2306.01048}{{\ttfamily 2306.01048}}].

\bibitem{Carenza:2023lci}
P.~Carenza, \emph{{Axion emission from supernovae: a cheatsheet}}, \href{https://doi.org/10.1140/epjp/s13360-023-04484-2}{\emph{Eur. Phys. J. Plus} {\bfseries 138} (2023) 836} [\href{https://arxiv.org/abs/2309.14798}{{\ttfamily 2309.14798}}].

\bibitem{Arvanitaki:2014dfa}
A.~Arvanitaki and A.~A. Geraci, \emph{{Resonantly Detecting Axion-Mediated Forces with Nuclear Magnetic Resonance}}, \href{https://doi.org/10.1103/PhysRevLett.113.161801}{\emph{Phys. Rev. Lett.} {\bfseries 113} (2014) 161801} [\href{https://arxiv.org/abs/1403.1290}{{\ttfamily 1403.1290}}].

\bibitem{Crescini:2017uxs}
N.~Crescini, C.~Braggio, G.~Carugno, P.~Falferi, A.~Ortolan and G.~Ruoso, \emph{{Improved constraints on monopole-dipole interaction mediated by pseudo-scalar bosons}}, \href{https://doi.org/10.1016/j.physletb.2017.09.019}{\emph{Phys. Lett. B} {\bfseries 773} (2017) 677} [\href{https://arxiv.org/abs/1705.06044}{{\ttfamily 1705.06044}}].

\bibitem{JacksonKimball:2017elr}
D.~F. Jackson~Kimball et~al., \emph{{Overview of the Cosmic Axion Spin Precession Experiment (CASPEr)}}, \href{https://doi.org/10.1007/978-3-030-43761-9_13}{\emph{Springer Proc. Phys.} {\bfseries 245} (2020) 105} [\href{https://arxiv.org/abs/1711.08999}{{\ttfamily 1711.08999}}].

\bibitem{Georgi:1986kr}
H.~Georgi and L.~Randall, \emph{{Flavor Conserving CP Violation in Invisible Axion Models}}, \href{https://doi.org/10.1016/0550-3213(86)90022-2}{\emph{Nucl. Phys. B} {\bfseries 276} (1986) 241}.

\bibitem{Georgi:1986df}
H.~Georgi, D.~B. Kaplan and L.~Randall, \emph{{Manifesting the Invisible Axion at Low-energies}}, \href{https://doi.org/10.1016/0370-2693(86)90688-X}{\emph{Phys. Lett. B} {\bfseries 169} (1986) 73}.

\bibitem{Peccei:1988ci}
R.~D. Peccei, \emph{{The Strong {CP} Problem}}, \href{https://doi.org/10.1142/9789814503280_0013}{\emph{Adv. Ser. Direct. High Energy Phys.} {\bfseries 3} (1989) 503}.

\bibitem{GrillidiCortona:2015jxo}
G.~Grilli~di Cortona, E.~Hardy, J.~Pardo~Vega and G.~Villadoro, \emph{{The QCD axion, precisely}}, \href{https://doi.org/10.1007/JHEP01(2016)034}{\emph{JHEP} {\bfseries 01} (2016) 034} [\href{https://arxiv.org/abs/1511.02867}{{\ttfamily 1511.02867}}].

\bibitem{Choi:2021ign}
K.~Choi, H.~J. Kim, H.~Seong and C.~S. Shin, \emph{{Axion emission from supernova with axion-pion-nucleon contact interaction}}, \href{https://doi.org/10.1007/JHEP02(2022)143}{\emph{JHEP} {\bfseries 02} (2022) 143} [\href{https://arxiv.org/abs/2110.01972}{{\ttfamily 2110.01972}}].

\bibitem{Ho:2022oaw}
S.-Y. Ho, J.~Kim, P.~Ko and J.-h. Park, \emph{{Supernova axion emissivity with \ensuremath{\Delta}(1232) resonance in heavy baryon chiral perturbation theory}}, \href{https://doi.org/10.1103/PhysRevD.107.075002}{\emph{Phys. Rev. D} {\bfseries 107} (2023) 075002} [\href{https://arxiv.org/abs/2212.01155}{{\ttfamily 2212.01155}}].

\bibitem{ParticleDataGroup:2022pth}
{\scshape Particle Data Group} Collaboration, R.~L. Workman et~al., \emph{{Review of Particle Physics}}, \href{https://doi.org/10.1093/ptep/ptac097}{\emph{PTEP} {\bfseries 2022} (2022) 083C01}.

\bibitem{Bauer:2017ris}
M.~Bauer, M.~Neubert and A.~Thamm, \emph{{Collider Probes of Axion-Like Particles}}, \href{https://doi.org/10.1007/JHEP12(2017)044}{\emph{JHEP} {\bfseries 12} (2017) 044} [\href{https://arxiv.org/abs/1708.00443}{{\ttfamily 1708.00443}}].

\bibitem{Bauer:2021mvw}
M.~Bauer, M.~Neubert, S.~Renner, M.~Schnubel and A.~Thamm, \emph{{Flavor probes of axion-like particles}}, \href{https://doi.org/10.1007/JHEP09(2022)056}{\emph{JHEP} {\bfseries 09} (2022) 056} [\href{https://arxiv.org/abs/2110.10698}{{\ttfamily 2110.10698}}].

\bibitem{Bauer:2020jbp}
M.~Bauer, M.~Neubert, S.~Renner, M.~Schnubel and A.~Thamm, \emph{{The Low-Energy Effective Theory of Axions and ALPs}}, \href{https://doi.org/10.1007/JHEP04(2021)063}{\emph{JHEP} {\bfseries 04} (2021) 063} [\href{https://arxiv.org/abs/2012.12272}{{\ttfamily 2012.12272}}].

\bibitem{deDivitiis:2013xla}
{\scshape RM123} Collaboration, G.~M. de~Divitiis, R.~Frezzotti, V.~Lubicz, G.~Martinelli, R.~Petronzio, G.~C. Rossi, F.~Sanfilippo, S.~Simula and N.~Tantalo, \emph{{Leading isospin breaking effects on the lattice}}, \href{https://doi.org/10.1103/PhysRevD.87.114505}{\emph{Phys. Rev. D} {\bfseries 87} (2013) 114505} [\href{https://arxiv.org/abs/1303.4896}{{\ttfamily 1303.4896}}].

\bibitem{MILC:2015ypt}
{\scshape MILC} Collaboration, S.~Basak et~al., \emph{{Electromagnetic effects on the light hadron spectrum}}, \href{https://doi.org/10.1088/1742-6596/640/1/012052}{\emph{J. Phys. Conf. Ser.} {\bfseries 640} (2015) 012052} [\href{https://arxiv.org/abs/1510.04997}{{\ttfamily 1510.04997}}].

\bibitem{Horsley:2015eaa}
R.~Horsley et~al., \emph{{Isospin splittings of meson and baryon masses from three-flavor lattice QCD + QED}}, \href{https://doi.org/10.1088/0954-3899/43/10/10LT02}{\emph{J. Phys. G} {\bfseries 43} (2016) 10LT02} [\href{https://arxiv.org/abs/1508.06401}{{\ttfamily 1508.06401}}].

\bibitem{Gavela:2023tzu}
B.~Gavela, P.~Qu\'\i{}lez and M.~Ramos, \emph{{The QCD axion sum rule}}, \href{https://doi.org/10.1007/JHEP04(2024)056}{\emph{JHEP} {\bfseries 04} (2024) 056} [\href{https://arxiv.org/abs/2305.15465}{{\ttfamily 2305.15465}}].

\bibitem{DiLuzio:2024jip}
L.~Di~Luzio, A.~W.~M. Guerrera, X.~Ponce~D\'\i{}az and S.~Rigolin, \emph{{Axion-Like Particles in Radiative Quarkonia Decays}},  \href{https://arxiv.org/abs/2402.12454}{{\ttfamily 2402.12454}}.

\bibitem{Carena:1988kr}
M.~Carena and R.~D. Peccei, \emph{{The Effective Lagrangian for Axion Emission From {SN1987A}}}, \href{https://doi.org/10.1103/PhysRevD.40.652}{\emph{Phys. Rev. D} {\bfseries 40} (1989) 652}.

\bibitem{Brinkmann:1988vi}
R.~P. Brinkmann and M.~S. Turner, \emph{{Numerical Rates for Nucleon-Nucleon Axion Bremsstrahlung}}, \href{https://doi.org/10.1103/PhysRevD.38.2338}{\emph{Phys. Rev. D} {\bfseries 38} (1988) 2338}.

\bibitem{Raffelt:1993ix}
G.~Raffelt and D.~Seckel, \emph{{A selfconsistent approach to neutral current processes in supernova cores}}, \href{https://doi.org/10.1103/PhysRevD.52.1780}{\emph{Phys. Rev. D} {\bfseries 52} (1995) 1780} [\href{https://arxiv.org/abs/astro-ph/9312019}{{\ttfamily astro-ph/9312019}}].

\bibitem{1996slfp.book.....R}
G.~G. {Raffelt}, \emph{{Stars as laboratories for fundamental physics : the astrophysics of neutrinos, axions, and other weakly interacting particles}}. 1996.

\bibitem{Turner:1991ax}
M.~S. Turner, \emph{{Dirac neutrinos and SN1987A}}, \href{https://doi.org/10.1103/PhysRevD.45.1066}{\emph{Phys. Rev. D} {\bfseries 45} (1992) 1066}.

\bibitem{Keil:1996ju}
W.~Keil, H.-T. Janka, D.~N. Schramm, G.~Sigl, M.~S. Turner and J.~R. Ellis, \emph{{A Fresh look at axions and SN-1987A}}, \href{https://doi.org/10.1103/PhysRevD.56.2419}{\emph{Phys. Rev. D} {\bfseries 56} (1997) 2419} [\href{https://arxiv.org/abs/astro-ph/9612222}{{\ttfamily astro-ph/9612222}}].

\bibitem{Ericson:1988wr}
T.~E.~O. Ericson and J.~F. Mathiot, \emph{{Axion Emission from SN 1987a: Nuclear Physics Constraints}}, \href{https://doi.org/10.1016/0370-2693(89)91103-9}{\emph{Phys. Lett. B} {\bfseries 219} (1989) 507}.

\bibitem{Raffelt:1991pw}
G.~Raffelt and D.~Seckel, \emph{{Multiple scattering suppression of the bremsstrahlung emission of neutrinos and axions in supernovae}}, \href{https://doi.org/10.1103/PhysRevLett.67.2605}{\emph{Phys. Rev. Lett.} {\bfseries 67} (1991) 2605}.

\bibitem{Janka:1995ir}
H.-T. Janka, W.~Keil, G.~Raffelt and D.~Seckel, \emph{{Nucleon spin fluctuations and the supernova emission of neutrinos and axions}}, \href{https://doi.org/10.1103/PhysRevLett.76.2621}{\emph{Phys. Rev. Lett.} {\bfseries 76} (1996) 2621} [\href{https://arxiv.org/abs/astro-ph/9507023}{{\ttfamily astro-ph/9507023}}].

\bibitem{Carenza:2020cis}
P.~Carenza, B.~Fore, M.~Giannotti, A.~Mirizzi and S.~Reddy, \emph{{Enhanced Supernova Axion Emission and its Implications}}, \href{https://doi.org/10.1103/PhysRevLett.126.071102}{\emph{Phys. Rev. Lett.} {\bfseries 126} (2021) 071102} [\href{https://arxiv.org/abs/2010.02943}{{\ttfamily 2010.02943}}].

\bibitem{Fore:2019wib}
B.~Fore and S.~Reddy, \emph{{Pions in hot dense matter and their astrophysical implications}}, \href{https://doi.org/10.1103/PhysRevC.101.035809}{\emph{Phys. Rev. C} {\bfseries 101} (2020) 035809} [\href{https://arxiv.org/abs/1911.02632}{{\ttfamily 1911.02632}}].

\bibitem{SNarchive}
\emph{{Garching core-collapse supernova research archive}},  \url{https://wwwmpa.mpa-garching.mpg.de/ccsnarchive//}.

\bibitem{Sukhbold:2017cnt}
T.~Sukhbold, S.~Woosley and A.~Heger, \emph{{A High-resolution Study of Presupernova Core Structure}}, \href{https://doi.org/10.3847/1538-4357/aac2da}{\emph{Astrophys. J.} {\bfseries 860} (2018) 93} [\href{https://arxiv.org/abs/1710.03243}{{\ttfamily 1710.03243}}].

\bibitem{Rampp:2002bq}
M.~Rampp and H.~T. Janka, \emph{{Radiation hydrodynamics with neutrinos: Variable Eddington factor method for core collapse supernova simulations}}, \href{https://doi.org/10.1051/0004-6361:20021398}{\emph{Astron. Astrophys.} {\bfseries 396} (2002) 361} [\href{https://arxiv.org/abs/astro-ph/0203101}{{\ttfamily astro-ph/0203101}}].

\bibitem{2020PhRvL.125e1104B}
R.~{Bollig}, W.~{DeRocco}, P.~W. {Graham} and H.-T. {Janka}, \emph{{Muons in Supernovae: Implications for the Axion-Muon Coupling}}, \href{https://doi.org/10.1103/PhysRevLett.125.051104}{\emph{\prl} {\bfseries 125} (2020) 051104} [\href{https://arxiv.org/abs/2005.07141}{{\ttfamily 2005.07141}}].

\bibitem{Caputo:2021rux}
A.~Caputo, G.~Raffelt and E.~Vitagliano, \emph{{Muonic boson limits: Supernova redux}}, \href{https://doi.org/10.1103/PhysRevD.105.035022}{\emph{Phys. Rev. D} {\bfseries 105} (2022) 035022} [\href{https://arxiv.org/abs/2109.03244}{{\ttfamily 2109.03244}}].

\bibitem{Caputo:2022mah}
A.~Caputo, H.-T. Janka, G.~Raffelt and E.~Vitagliano, \emph{{Low-Energy Supernovae Severely Constrain Radiative Particle Decays}}, \href{https://doi.org/10.1103/PhysRevLett.128.221103}{\emph{Phys. Rev. Lett.} {\bfseries 128} (2022) 221103} [\href{https://arxiv.org/abs/2201.09890}{{\ttfamily 2201.09890}}].

\bibitem{Fiorillo:2023frv}
D.~F.~G. Fiorillo, M.~Heinlein, H.-T. Janka, G.~Raffelt, E.~Vitagliano and R.~Bollig, \emph{{Supernova simulations confront SN 1987A neutrinos}}, \href{https://doi.org/10.1103/PhysRevD.108.083040}{\emph{Phys. Rev. D} {\bfseries 108} (2023) 083040} [\href{https://arxiv.org/abs/2308.01403}{{\ttfamily 2308.01403}}].

\bibitem{Fore:2023gwv}
B.~Fore, N.~Kaiser, S.~Reddy and N.~C. Warrington, \emph{{The mass of charged pions in neutron star matter}},  \href{https://arxiv.org/abs/2301.07226}{{\ttfamily 2301.07226}}.

\bibitem{Lucente:2020whw}
G.~Lucente, P.~Carenza, T.~Fischer, M.~Giannotti and A.~Mirizzi, \emph{{Heavy axion-like particles and core-collapse supernovae: constraints and impact on the explosion mechanism}}, \href{https://doi.org/10.1088/1475-7516/2020/12/008}{\emph{JCAP} {\bfseries 12} (2020) 008} [\href{https://arxiv.org/abs/2008.04918}{{\ttfamily 2008.04918}}].

\bibitem{Raffelt:2006rj}
G.~G. Raffelt, \emph{{Axions: Motivation, limits and searches}}, \href{https://doi.org/10.1088/1751-8113/40/25/S05}{\emph{J. Phys. A} {\bfseries 40} (2007) 6607} [\href{https://arxiv.org/abs/hep-ph/0611118}{{\ttfamily hep-ph/0611118}}].

\bibitem{Calore:2021klc}
F.~Calore, P.~Carenza, M.~Giannotti, J.~Jaeckel, G.~Lucente and A.~Mirizzi, \emph{{Supernova bounds on axionlike particles coupled with nucleons and electrons}}, \href{https://doi.org/10.1103/PhysRevD.104.043016}{\emph{Phys. Rev. D} {\bfseries 104} (2021) 043016} [\href{https://arxiv.org/abs/2107.02186}{{\ttfamily 2107.02186}}].

\bibitem{Jaeckel:2017tud}
J.~Jaeckel, P.~C. Malta and J.~Redondo, \emph{{Decay photons from the axionlike particles burst of type II supernovae}}, \href{https://doi.org/10.1103/PhysRevD.98.055032}{\emph{Phys. Rev. D} {\bfseries 98} (2018) 055032} [\href{https://arxiv.org/abs/1702.02964}{{\ttfamily 1702.02964}}].

\bibitem{Altmann:1995bw}
M.~Altmann, F.~von Feilitzsch, C.~Hagner, L.~Oberauer, Y.~Declais and E.~Kajfasz, \emph{{Search for the electron positron decay of axions and axion - like particles at a nuclear power reactor at Bugey}}, \href{https://doi.org/10.1007/BF01566670}{\emph{Z. Phys. C} {\bfseries 68} (1995) 221}.

\bibitem{Falk:1978kf}
S.~W. Falk and D.~N. Schramm, \emph{{Limits From Supernovae on Neutrino Radiative Lifetimes}}, \href{https://doi.org/10.1016/0370-2693(78)90417-3}{\emph{Phys. Lett. B} {\bfseries 79} (1978) 511}.

\bibitem{Sung:2019xie}
A.~Sung, H.~Tu and M.-R. Wu, \emph{{New constraint from supernova explosions on light particles beyond the Standard Model}}, \href{https://doi.org/10.1103/PhysRevD.99.121305}{\emph{Phys. Rev. D} {\bfseries 99} (2019) 121305} [\href{https://arxiv.org/abs/1903.07923}{{\ttfamily 1903.07923}}].

\bibitem{Stockinger:2020hse}
G.~Stockinger et~al., \emph{{Three-dimensional Models of Core-collapse Supernovae From Low-mass Progenitors With Implications for Crab}}, \href{https://doi.org/10.1093/mnras/staa1691}{\emph{Mon. Not. Roy. Astron. Soc.} {\bfseries 496} (2020) 2039} [\href{https://arxiv.org/abs/2005.02420}{{\ttfamily 2005.02420}}].

\bibitem{Yang:2015ooa}
H.~Yang and R.~A. Chevalier, \emph{{Evolution of the Crab nebula in a low energy supernova}}, \href{https://doi.org/10.1088/0004-637X/806/2/153}{\emph{Astrophys. J.} {\bfseries 806} (2015) 153} [\href{https://arxiv.org/abs/1505.03211}{{\ttfamily 1505.03211}}].

\bibitem{Ouchi:2017cza}
R.~Ouchi and K.~Maeda, \emph{{Radii and Mass-loss Rates of Type IIb Supernova Progenitors}}, \href{https://doi.org/10.3847/1538-4357/aa6ea9}{\emph{Astrophys. J.} {\bfseries 840} (2017) 90} [\href{https://arxiv.org/abs/1705.02430}{{\ttfamily 1705.02430}}].

\bibitem{Goldberg:2020czt}
J.~A. Goldberg and L.~Bildsten, \emph{{The Value of Progenitor Radius Measurements for Explosion Modeling of Type II-Plateau Supernovae}}, \href{https://doi.org/10.3847/2041-8213/ab9300}{\emph{Astrophys. J. Lett.} {\bfseries 895} (2020) L45} [\href{https://arxiv.org/abs/2005.07290}{{\ttfamily 2005.07290}}].

\bibitem{Langhoff:2022bij}
K.~Langhoff, N.~J. Outmezguine and N.~L. Rodd, \emph{{Irreducible Axion Background}}, \href{https://doi.org/10.1103/PhysRevLett.129.241101}{\emph{Phys. Rev. Lett.} {\bfseries 129} (2022) 241101} [\href{https://arxiv.org/abs/2209.06216}{{\ttfamily 2209.06216}}].

\bibitem{Muller:2023pip}
E.~M\"uller, P.~Carenza, C.~Eckner and A.~Goobar, \emph{{Constraining MeV-scale axionlike particles with Fermi-LAT observations of SN 2023ixf}}, \href{https://doi.org/10.1103/PhysRevD.109.023018}{\emph{Phys. Rev. D} {\bfseries 109} (2024) 023018} [\href{https://arxiv.org/abs/2306.16397}{{\ttfamily 2306.16397}}].

\bibitem{Chupp:1989kx}
E.~L. Chupp, W.~T. Vestrand and C.~Reppin, \emph{{Experimental Limits on the Radiative Decay of {SN1987A} Neutrinos}}, \href{https://doi.org/10.1103/PhysRevLett.62.505}{\emph{Phys. Rev. Lett.} {\bfseries 62} (1989) 505}.

\bibitem{Oberauer:1993yr}
L.~Oberauer, C.~Hagner, G.~Raffelt and E.~Rieger, \emph{{Supernova bounds on neutrino radiative decays}}, \href{https://doi.org/10.1016/0927-6505(93)90004-W}{\emph{Astropart. Phys.} {\bfseries 1} (1993) 377}.

\bibitem{Hoof:2022xbe}
S.~Hoof and L.~Schulz, \emph{{Updated constraints on axion-like particles from temporal information in supernova SN1987A gamma-ray data}}, \href{https://doi.org/10.1088/1475-7516/2023/03/054}{\emph{JCAP} {\bfseries 03} (2023) 054} [\href{https://arxiv.org/abs/2212.09764}{{\ttfamily 2212.09764}}].

\bibitem{Muller:2023vjm}
E.~M\"uller, F.~Calore, P.~Carenza, C.~Eckner and M.~C.~D. Marsh, \emph{{Investigating the gamma-ray burst from decaying MeV-scale axion-like particles produced in supernova explosions}}, \href{https://doi.org/10.1088/1475-7516/2023/07/056}{\emph{JCAP} {\bfseries 07} (2023) 056} [\href{https://arxiv.org/abs/2304.01060}{{\ttfamily 2304.01060}}].

\bibitem{Jaffe:1995sw}
A.~H. Jaffe and M.~S. Turner, \emph{{Gamma-rays and the decay of neutrinos from SN1987A}}, \href{https://doi.org/10.1103/PhysRevD.55.7951}{\emph{Phys. Rev. D} {\bfseries 55} (1997) 7951} [\href{https://arxiv.org/abs/astro-ph/9601104}{{\ttfamily astro-ph/9601104}}].

\bibitem{Diamond:2023scc}
M.~Diamond, D.~F.~G. Fiorillo, G.~Marques-Tavares and E.~Vitagliano, \emph{{Axion-sourced fireballs from supernovae}}, \href{https://doi.org/10.1103/PhysRevD.107.103029}{\emph{Phys. Rev. D} {\bfseries 107} (2023) 103029} [\href{https://arxiv.org/abs/2303.11395}{{\ttfamily 2303.11395}}]. [Erratum: Phys.Rev.D 108, 049902 (2023)].

\bibitem{Raffelt:2011ft}
G.~G. Raffelt, J.~Redondo and N.~Viaux~Maira, \emph{{The meV mass frontier of axion physics}}, \href{https://doi.org/10.1103/PhysRevD.84.103008}{\emph{Phys. Rev. D} {\bfseries 84} (2011) 103008} [\href{https://arxiv.org/abs/1110.6397}{{\ttfamily 1110.6397}}].

\bibitem{Beacom:2010kk}
J.~F. Beacom, \emph{{The Diffuse Supernova Neutrino Background}}, \href{https://doi.org/10.1146/annurev.nucl.010909.083331}{\emph{Ann. Rev. Nucl. Part. Sci.} {\bfseries 60} (2010) 439} [\href{https://arxiv.org/abs/1004.3311}{{\ttfamily 1004.3311}}].

\bibitem{Eckner:2021kjb}
C.~Eckner, F.~Calore, P.~Carenza, M.~Giannotti, J.~Jaeckel, F.~Sivo and A.~Mirizzi, \emph{{Constraining the diffuse supernova axion-like-particle background with high-latitude Fermi-LAT data}}, \href{https://doi.org/10.22323/1.395.0543}{\emph{PoS} {\bfseries ICRC2021} (2021) 543}.

\bibitem{Calore:2021hhn}
F.~Calore, P.~Carenza, C.~Eckner, T.~Fischer, M.~Giannotti, J.~Jaeckel, K.~Kotake, T.~Kuroda, A.~Mirizzi and F.~Sivo, \emph{{3D template-based Fermi-LAT constraints on the diffuse supernova axion-like particle background}}, \href{https://doi.org/10.1103/PhysRevD.105.063028}{\emph{Phys. Rev. D} {\bfseries 105} (2022) 063028} [\href{https://arxiv.org/abs/2110.03679}{{\ttfamily 2110.03679}}].

\bibitem{Calore:2020tjw}
F.~Calore, P.~Carenza, M.~Giannotti, J.~Jaeckel and A.~Mirizzi, \emph{{Bounds on axionlike particles from the diffuse supernova flux}}, \href{https://doi.org/10.1103/PhysRevD.102.123005}{\emph{Phys. Rev. D} {\bfseries 102} (2020) 123005} [\href{https://arxiv.org/abs/2008.11741}{{\ttfamily 2008.11741}}].

\bibitem{Priya:2017bmm}
A.~Priya and C.~Lunardini, \emph{{Diffuse neutrinos from luminous and dark supernovae: prospects for upcoming detectors at the $O$(10) kt scale}}, \href{https://doi.org/10.1088/1475-7516/2017/11/031}{\emph{JCAP} {\bfseries 11} (2017) 031} [\href{https://arxiv.org/abs/1705.02122}{{\ttfamily 1705.02122}}].

\bibitem{2008ICRC....3.1085S}
J.~G. {Stacy}, W.~{Collmar}, A.~{Strong}, V.~{Schonfelder} and A.~{Carraminana}, \emph{{Limits on MeV Gamma-Ray Emission from Active Galaxies and Other Unidentified High-Latitude Gamma-Ray Sources Observed with COMPTEL}},  in \emph{International Cosmic Ray Conference}, vol.~3 of \emph{International Cosmic Ray Conference}, pp.~1085--1088, Jan., 2008.

\end{thebibliography}\endgroup

\clearpage
\onecolumngrid

\appendix
\section{Interplay between an ALP coupled to QCD and the QCD axion}

\label{sec:axionAndALP}

In the main part of the paper we consider only an ALP coupled to QCD. In this case we leave the solution of the strong CP problem unspecified. In principle, one can wonder how the picture changes when a solution of this problem, in the form of a QCD axion, is introduced in addition to the ALP. In this scenario, the Lagrangian in Eq.~\eqref{eq:LaaboveQCD} would acquire an extra term
\begin{equation}
\mathcal{L}= \frac{g^{2}}{32\pi^{2}f_{a}}\left(a_{\rm QCD} +c_{g}a\right)  
G^a_{\mu\nu} \tilde G^{a\mu\nu} +\frac{(m_{a,0})^{2}}{2}(a-a_{0})^{2}\,,
\label{eq:QCDAx+ALP}
\end{equation}
where we neglect the quark couplings for simplicity and have absorbed any possible difference between the two decay constants of $a_{\rm QCD}$ and $a$ into $c_g$. Note that, at a fundamental level, the constant $a_{0}$ can always be absorbed by a shift of the ALP field, $a\to a+a_{0}$. This shift needs to be appropriately compensated by the QCD axion cancelling a $\theta$-term, i.e. $a_{\rm QCD}\to a_{\rm QCD}-c_{g}a_{0}$. Thus, in the following we consider $a_{0}=0$ in full generality. This conclusion may not be valid in case of a generic potential for the ALP.
It should be noted that $a_{\rm QCD}$ does not have a bare mass term, as it would either be negligible or spoil the solution to the strong CP problem. However, after confinement, both axion and ALP gain a mass through non-perturbative QCD dynamics, leading to a non-diagonal mass matrix~\cite{Gavela:2023tzu}
\begin{equation}
  \mathbf{M}^{2}=m_{\rm QCD}^{2}  \begin{pmatrix}
        1&c_{g}\\
        c_{g}&c_{g}^{2}+\frac{m_{a,0}^{2}}{m_{\rm QCD}^{2}}
    \end{pmatrix}\,,
\end{equation}
where $m_{\rm QCD}^{2}$ is the QCD-induced mass. Therefore, the mass eigenvalues read
\begin{equation}
    m_{1,2}^2=\frac{1}{2}\,\left( m_{a,0}^2+C_G^2\,m_{\rm QCD}^2 \pm \Delta m^2\right)\,,
    \label{eq:eigenvalues}
\end{equation}
where we have defined
\begin{equation}
    \begin{split}
        &\Delta m^2=\sqrt{(m_{a,0}^2+C_G^2\,m_{\rm QCD}^2)^2-4m_{a,0}^2\,m_{\rm QCD}^2}\\
        &C_G^2=1+c_g^2\,.
    \end{split}
\end{equation}
Remarkably, in the limit $m_{a,0}\gg m_{\rm QCD}$, which is the physical case considered in this work, the mass eigenvalues reduce to $m_1\simeq m_{a,0}$ and $m_2\simeq c_g^2 m_{\rm QCD}$. In the same way, the corresponding eigenstates 
\begin{equation}
    \begin{split}
    & |a_{1}\rangle=   \begin{pmatrix}
        -\frac{m_{a,0}^{2}+(c_{g}^{2}-1)m_{\rm QCD}^{2}-\Delta m^{2}}{2c_{g} m_{\rm QCD}^{2}}
        \\
        1
        \end{pmatrix}\rightarrow \,|a_{\rm QCD}\rangle=   \begin{pmatrix}
        1
        \\
        0
        \end{pmatrix}\\
         & |a_{2}\rangle=   \begin{pmatrix}
        -\frac{m_{a,0}^{2}+(c_{g}^{2}-1)m_{\rm QCD}^{2}+\Delta m^{2}}{2c_{g} m_{\rm QCD}^{2}}
        \\
        1
        \end{pmatrix}\rightarrow \,|a\rangle=   \begin{pmatrix}
        0
        \\
        1
        \end{pmatrix}\,,\\
    \end{split}
\end{equation}
where we have employed a proper normalization.\\
These results suggest that the large mass splitting, $m_{a,0}\gg m_{\rm QCD}$,  leads to a negligible mixing between the two mass eigenstates, which are actually the eigenstates propagating in vacuum. Therefore, we can always look at the phenomenology of the ALP without referring to the QCD axion. Its existence might only strengthen the SN cooling bound, where the QCD axion is produced and escapes the SN core without giving observational signatures. Otherwise, it might be that the QCD axion is just weakly produced in the SN and, in this case, the cooling bound is unaffected.

\vspace{0.2cm}
The axion-ALP Lagrangian in Eq.~\eqref{eq:QCDAx+ALP} can be also written in terms of terms of
\begin{equation}
    \begin{split}
        a_{G\Tilde{G}}&=C_G^{-1}\,(a_{\rm QCD}+c_g\,a)\\
        a_{\perp}&=C_G^{-1}\,(c_g\,a_{\rm QCD}-\,a)\,.
    \end{split}
\end{equation}
It is worthy to highlight that, by employing this orthonormal basis, the only field coupling to the gluon field is $ a_{G\Tilde{G}}$. Thus, in this formalism, $a_{G\Tilde{G}}$ plays the role of a ``QCD ALP" as interaction eigenstate \footnote{ But only its $a_{\rm QCD}$ component solves the strong CP problem.}, while its orthogonal counterpart $a_{\perp}$ is decoupled from $G\Tilde{G}$ and appears just in the mass term
\begin{equation}
    \mathcal{L}= C_G\,\frac{g^{2}}{32\pi^{2}f_{a}}\,a_{G\Tilde{G}}  
    G^a_{\mu\nu} \tilde G^{a\mu\nu} -\frac{1}{2}\frac{(m_{a,0})^{2}}{C_G^2}\left(c_g\,{a_{G\Tilde{G}}-a_{\perp}}\right)^{2}\,.
    \label{eq:lagpic}
\end{equation}
In terms of this basis, the mass matrix after the QCD phase transition reads 
\begin{equation}
     \mathbf{M'}^{2}=\begin{pmatrix}
        C_G^2\,m_{\rm QCD}^2+\frac{c_g^2\,m_{a,O}^2}{C_G^2}&\,\,\,-\frac{c_g\,m_{a,O}^2}{C_G^2}\\
        -\frac{c_g\,m_{a,O}^2}{C_G^2}&-\frac{m_{a,O}^2}{C_G^2}\,.
    \end{pmatrix}\,,
\end{equation}
showing, as expected, the same eigenvalues as in Eq.~\eqref{eq:eigenvalues}. 
The corresponding eigenstates in the $m_{a,0}\gg m_{\rm QCD}$ limit can be written as 

\begin{equation}
    \begin{split}
    & |a'_{1}\rangle=   \begin{pmatrix}
        -\frac{(c_g^2-1) m_{a,0}^{2}+C_G^2(C_G^2 m_{\rm QCD}^{2}-\Delta m^{2})}{2c_{g} m_{\rm QCD}^{2}}
        \\
        1
        \end{pmatrix}\rightarrow \,|a_{\rm QCD}\rangle=  C_G^{-1} \begin{pmatrix}
        1
        \\
        c_g
        \end{pmatrix}\\
         & |a'_{2}\rangle=   \begin{pmatrix}
        -\frac{(c_g^2-1) m_{a,0}^{2}+C_G^2(C_G^2 m_{\rm QCD}^{2}+\Delta m^{2})}{2c_{g} m_{\rm QCD}^{2}}
        \\
        1
        \end{pmatrix}\rightarrow \,|a\rangle=  C_G^{-1} \begin{pmatrix}
        c_g
        \\
        -1
        \end{pmatrix}\,,\\
    \end{split}
\end{equation}
and the associated phenomenology reduces to the physics previously discussed. However, this formalism is useful, since it shows that a ``QCD ALP'' coupled to QCD only, with a non-diagonal mass term, has the same physical effects of a mixture of two axions: a massive ALP and a massless QCD axion.

The picture described by Eq.~\eqref{eq:lagpic} is convenient to recast known results on the QCD axion, to our QCD ALP. Namely, the low-energy interactions between the ALP and hadrons will be described in terms of Eq.~\eqref{eq:LaaboveQCD}, with the replacement $c_g\,a\to C_{G}\,a_{G\tilde{G}}$. Similarly for the ALP-photon interaction. This is the reason why the interactions in Eqs.~\eqref{eq:coupl}-\eqref{eq:C_gamma}, referring to the QCD ALP, feature just a rescaling in terms of $c_{g}$.

\end{document}